\documentclass[useAMS,usenatbib]{mn2e}

\usepackage{graphicx}\usepackage{epsfig}\usepackage{epsf}
\usepackage{amsmath}\usepackage{amssymb}\usepackage{stfloats}
\title[MCA-1B in M33]{A new and unusual LBV-like outburst from a
  Wolf-Rayet star in the outskirts of M33}

\author[Smith et al.]{Nathan Smith$^{1}$\thanks{E-mail:
    nathans@as.arizona.edu}, Jennifer E.\ Andrews$^1$, Maxwell
  Moe$^1$, Peter Milne$^1$, Christopher \newauthor Bilinski$^1$,
  Charles D.\ Kilpatrick$^2$, Wen-Fai Fong$^{3,4}$, Carles
  Badenes$^5$, Alexei V.\ \newauthor Filippenko$^{6,7}$, Mansi
  Kasliwal$^8$, and Jeffrey M. Silverman$^9$ \\ $^{1}$Steward
  Observatory, University of Arizona, 933 N.\ Cherry Ave., Tucson, AZ
  85721, USA \\ $^2$Department of Astronomy and Astrophysics,
  University of California, Santa Cruz, CA 95064, USA
  \\ $^3$Department of Physics and Astronomy, Northwestern University,
  2145 Sheridan Road, Evanston, IL, USA \\ $^4$Center for
  Interdisciplinary Exploration and Research in Astrophysics,
  Northwestern University, 1800 Sherman Ave, Evanston, IL, USA
  \\ $^5$Department of Physics and Astronomy, and Pittsburgh Particle
  Physics, Astrophysics and Cosmology Center (PITT PACC), University
  of \\ Pittsburgh, 3941 O’Hara Street, Pittsburgh, PA 15260, USA
  \\ $^6$Department of Astronomy, University of California, Berkeley,
  CA 94720-3411, USA \\ $^7$Miller Senior Fellow, Miller Institute for
  Basic Research in Science, University of California, Berkeley, CA
  94720, USA \\ $^8$Division of Physics, Mathematics, and Astronomy,
  California Institute of Technology, Pasadena, CA 91125, USA
  \\ $^9$Samba TV, San Francisco, CA 94107, USA}

\begin{document}

\pagerange{\pageref{firstpage}--\pageref{lastpage}} \pubyear{2019}
\maketitle
\label{firstpage}

\begin{abstract}

  MCA-1B (also called UIT003) is a luminous hot star in the western
  outskirts of M33, classified over 20~yr ago with a spectral type of
  Ofpe/WN9 and identified then as a candidate luminous blue variable
  (LBV).  Palomar Transient Factory data reveal that this star
  brightened in 2010, with a light curve resembling that of the
  classic LBV star AF~And in M31.  Other Ofpe/WN9 stars have erupted
  as LBVs, but MCA-1B was unusual because it remained hot.  It showed
  a WN-type spectrum throughout its eruption, whereas LBVs usually get
  much cooler.  MCA-1B showed an almost four-fold increase in
  bolometric luminosity and a doubling of its radius, but its
  temperature stayed $\gtrsim$29~kK. As it faded, it shifted to even
  hotter temperatures, exhibiting a WN7/WN8-type spectrum, and
  doubling its wind speed.  MCA-1B is reminiscent of some supernova
  impostors, and its location resembles the isolated environment of
  SN~2009ip.  It is most similar to HD~5980 (in the Small Magellanic
  Cloud) and GR~290 (also in M33).  Whereas these two LBVs exhibited
  B-type spectra in eruption, MCA-1B is the first clear case where a
  Wolf-Rayet (WR) spectrum persisted at all times.  Together, MCA-1B,
  HD~5980, and GR~290 constitute a class of WN-type LBVs, distinct from
  S~Doradus LBVs.  They are most interesting in the context of LBVs at
  low metallicity, a possible post-LBV/WR transition in binaries, and
  as likely Type~Ibn supernova progenitors.

\end{abstract}


\section{INTRODUCTION}

Luminous blue variables (LBVs) are evolved massive stars that are the
brightest blue irregular variables in star-forming galaxies.
Originally recognised as the classic Hubble-Sandage variables in M31
and M33 \citep{hs53}, these were later grouped together with a diverse
collection of irregular variable stars and referred to as ``LBVs'' by
\citet{conti84}.  LBVs have the highest mass-loss rates of any stars,
and this can profoundly affect a massive star's late-time evolution
\citep{smith14}.  However, their role in the evolution of massive
stars in general is uncertain.  In particular, the central idea that
their mass loss is the gateway to H-poor Wolf-Rayet (WR) stars in
single-star evolution \citep{conti76,mc94,so06,groh14} is problematic
because LBVs appear to be massive blue stragglers \citep{st15}.
``Massive blue stragglers'' in this context means that they appear
overluminous or too young for single-star evolution, as compared to
surrounding stars; see \citet{st15} and \citet{mojgan17} for details.

Our current understanding of LBVs is in flux, and many established
ideas about LBVs have turned out to be incorrect or incomplete.  Since
LBVs are quite rare (only a handful are known in the Milky Way), many
ideas about LBVs have been shaped by detailed study of a few
prototypical objects and then imposed upon the class as a whole.
However, continued study of nearby LBVs and extraglactic transients
has shown that numerous hallmarks of LBVs seem to break down upon
closer examination, and LBVs are phenomenologically diverse (see
\citealt{smith17}, and references therein).

Particularly relevant to the current paper is one of the traditionally
adopted defining characteristics of LBVs: their S~Doradus outbursts
show a brightening at visible wavelengths as a result of temperature
shifts due to a cooler pseudophotosphere forming in the denser wind.
The pseudophotosphere was the presumed explanation for why LBVs, at
maximum brightness, reside along a constant-temperature strip
\citep{hd94}.  This was thought to be the explanation for a shift in
energy distribution from the ultraviolet (UV) to the optical, causing
a dramatic brightening at {\it constant} bolometric luminosity
\citep{hd94}.  However, the notion that the cool temperature is caused
by a pseudophotosphere due to increased mass loss, and the idea that
these temperature changes occur at constant $L_{\rm bol}$, both appear
to be incorrect.  Investigations with quantitative spectroscopy
\citep{dekoter96,lamers95,groh06,groh09a,groh09b,mehner17} disproved
the conjecture that S~Doradus brightening events are caused by opaque
winds, because the mass-loss rates are not high enough to make such
large pseudophotospheres.  Moreover, bolometric luminosities during
S~Doradus eruptions are actually not constant.  There are also some
well-studied exceptions to the traditional behaviour of cool
temperatures in LBV eruptions, most notably in the eruption of the
massive eclipsing binary HD~5980 in the Small Magellanic Cloud (SMC)
and Romano's star (GR~290) in M33, which are discussed below.  A
limiting factor in understanding LBVs is that there are few of them,
and there is considerable diversity even among this small group,
making it difficult to identify the physics of their instability more
generally.  Moreover, with revised distances from {\it Gaia} DR2, many
LBVs do not reside on the S~Doradus instability strip \citep{smith19},
as generally assumed.


LBV candidates are more numerous than confirmed LBVs \citep{massey07}
and may hold additional clues to their nature.  LBV candidates occupy
a similar part of the Hertzsprung-Russell (HR) diagram as quiescent
LBVs and have similar spectral properties, but have not been caught
demonstrating the tell-tale dramatic photometric variability.  Some
have resolved circumstellar nebulae that seem to indicate a past
LBV-like eruptive mass-loss episode \citep{stahl86}, and they are
often presumed to be dormant LBVs.

Chief among the LBV candidates are the Ofpe/WN9 stars, intermediate
between Of-type stars and WN stars \citep{walborn77,walborn82a,bw89}.
The Ofpe/WN9 stars were first recognised as a class based on about ten
objects in the Large Magellanic Cloud (LMC), but have also been found
in a number of nearby galaxies.  Ofpe/WN9 stars are luminous,
typically log($L/L_{\odot}$) = 5.5--6.5, implying high initial masses;
also, they have hot temperatures of roughly 29--30 kK, and wind speeds
of roughly 400 km s$^{-1}$ \citep{crowther95a}.  Their spectra are
similar to those of classical LBVs in their hot quiescent state, and
many of the LBV candidates with shell nebulae are Ofpe/WN9 stars.
There is actually a direct connection between Ofpe/WN9 stars and
confirmed LBVs.  Two of the prototypical LBVs that have classic
S~Doradus cycles --- AG Carinae in the Milky Way and R127 in the LMC
--- both appear as Ofpe/WN9 stars in their hot quiescent state
\citep{walborn82b,ws82,stahl83,stahl01}.  Following these examples,
one expects a hot Ofpe/WN9 star to become cooler (i.e., an F
supergiant) as it brightens if it has a 1--2 mag LBV eruption.  This
brings us to the variable Ofpe/WN9 star in M33 that we discuss here,
which flagrantly defies this expectation.


The first Ofpe/WN9 star identified in M33 was MCA-1B, which was found
accidentally \citep{willis92}.  It was identified spectroscopically
when it was observed instead of the intended target, the fainter
neighbouring WC star MCA-1 \citep{mca}, which is only about 2{\arcsec}
away.  Since then, a handful of additional Ofpe/WN9 stars were found
in M33 \citep{massey96,bianchi04}.

MCA-1B was studied in detail shortly after its initial discovery by
\citet{willis92}, and a quantitative spectroscopic analysis was
performed on high-resolution spectra \citep{crowther95,smith95}.  The
estimated physical parameters were a stellar temperature of 29 kK, a
bolometric luminosity of log($L/L_{\odot}$) = 5.8, a mass-loss rate of
$\dot{M} = 10^{-4}$ $M_{\odot}$ yr$^{-1}$, a wind terminal speed of
$v_{\infty}$ = 420 km s$^{-1}$, and a probable initial mass of around
50 $M_{\odot}$, based on its position on the HR diagram compared to
single-star evolutionary models.  Those authors noted its similarity
to other Ofpe/WN9 stars in the LMC, especially the classic Ofpe/WN9
star R84 (which was presumed to be a dormant LBV), and they concluded
that MCA-1B was likely to be a dormant LBV as well.  The star is one
of the brightest stars in M33 in the UV (it is also named UIT003), and
its UV spectrum obtained in 2000 December has been studied in detail by
\citet{bianchi04}.  Those authors derived somewhat different values in
their analysis of the UV/optical spectrum: $T_{\rm eff}$ = 30 kK,
log($L/L_{\odot}$) = 5.55, $\dot{M}$ = $2 \times 10^{-6}$
$M_{\odot}$ yr$^{-1}$, $v_{\infty}$ = 1000 km s$^{-1}$, and a likely
initial mass around 25 $M_{\odot}$.  Since these UV/optical
observations of MCA-1B were obtained several years after the spectrum
analyzed by \citet{smith95}, the different derived values may be a
consequence of real variability, but may also be attributable to a
somewhat lower reddening value adopted by \citet{bianchi04}.  Some
differences in $V$-band magnitude have been reported in the literature
for MCA-1B: $V=17.7$ mag in 1992 with 0$\farcs$8 seeing
\citep{smith95}, $V=17.43$ mag in 1993 with 1$\farcs$4--1$\farcs$7
seeing \citep{massey96}, and $V=17.63$ mag in 2000 with 0$\farcs$9
seeing \citep{massey06}.  Because MCA-1B is in a crowded region (this
is discussed later in the paper), however, it is plausible that these
differences are in part due to different amounts of contamination by
neighbouring stars (i.e., the dates with poorer seeing give brighter
magnitudes).  Significantly brighter $B$-band magnitudes were derived
from low-resolution ($12.8''$ mm$^{-1}$) photographic plates from
the 1980s \citep{kurtev99}, but the contamination from surrounding
stars is probably severe in those data. Thus, it remains uncertain if
there was significant photometric variability of MCA-1B before the
recent epoch reported here.

In this paper we discuss the identification by the Palomar Transient
Factory (PTF) of the variable source PTF10vyq in M33, which turns out
to be a nonperiodic brightening of the well-known Ofpe/WN9 star
MCA-1B.  The onset of its irregular variability suggests that MCA-1B
is in fact an LBV, and was previously therefore a dormant LBV as
\citet{crowther95} and \citet{smith95} suspected.  Upon examination,
however, the new outburst is peculiar compared with traditional
S~Doradus outbursts of LBVs.

MCA-1B/PTF10vyq is located in the western outskirts of M33, and is
therefore of additional interest, since some recent peculiar SNe
related to LBVs, such as SN~2009ip \citep{smith16}, are found in
environments that are surprisingly remote for the high luminosity and
high implied masses of their progenitors.  The remote location
combined with the metallicity gradient of M33 \citep{magrini07} also
suggests that MCA-1B may provide intriguing clues to LBV behaviour at
low metallicity.  Throughout, we adopt a distance to M33 of 830 kpc,
following \citet{massey13}.  This is between estimates of 800 kpc
\citep{m09,patel17} and 960 kpc \citep{bonanos06}; adopting either of
these these would only slightly alter the absolute magnitudes and
implications for the initial mass of the star.

In Section 2 we present the discovery and follow-up of PTF10vyq, which
is a brightening of the Ofpe/WN9 star MCA-1B.  We discuss its remote
location in Section 3, and crowding in its vicinity in Section 4. Its
physical parameters and their variability are presented in Section
5. We summarise the main results in Section 6.

\begin{figure*}
  \includegraphics[width=5.8in]{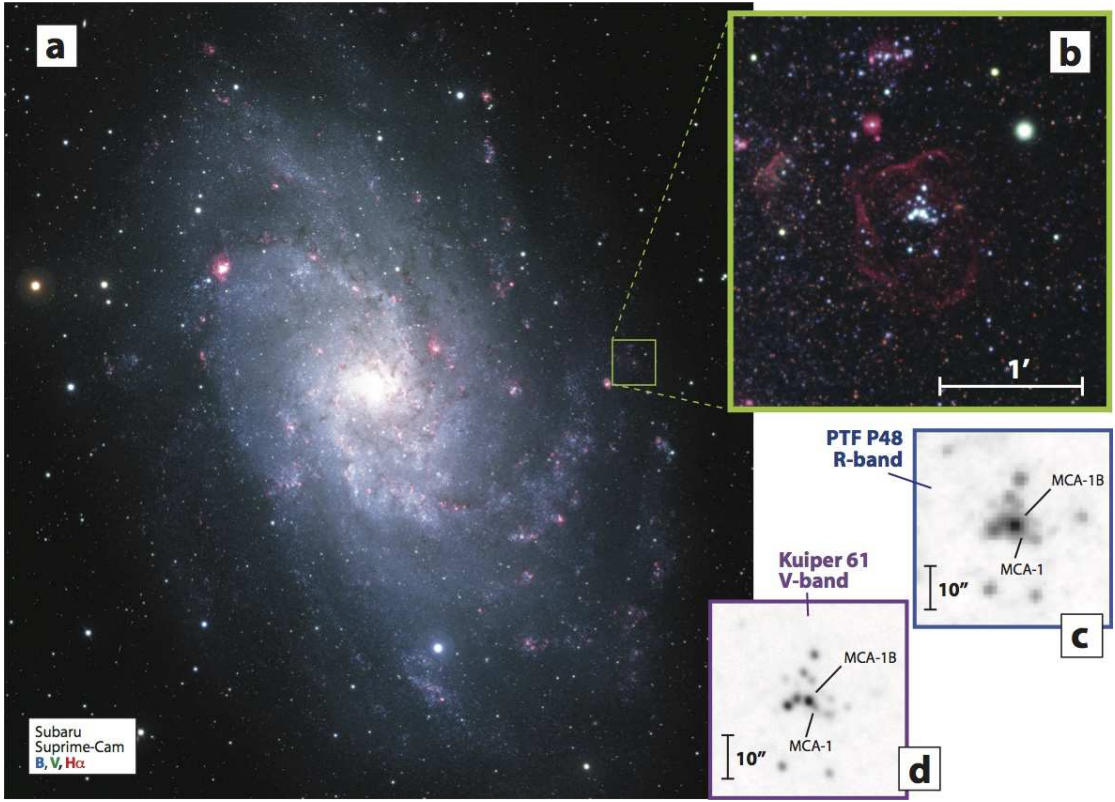}
\caption{(a) A wide-field colour image of M33 taken with Subaru
  Suprime-Cam, with blue = $B$, green = $V$, and red = H$\alpha$
  (image credit: National Astronomical Observatory of Japan).  North
  is up and east to the left.  The locations of MCA-1B and its cluster
  are in the green box.  (b) A close-up view of the same image showing
  the immediate environment around MCA-1B, with its host cluster and
  H~{\sc ii} region shell. (c) PTF P48 $R$-band image of the host
  cluster of MCA-1B.  (d) K61 $V$-band image of the cluster.}
\label{fig:img}
\end{figure*}

\begin{figure*}
  \includegraphics[width=5.0in]{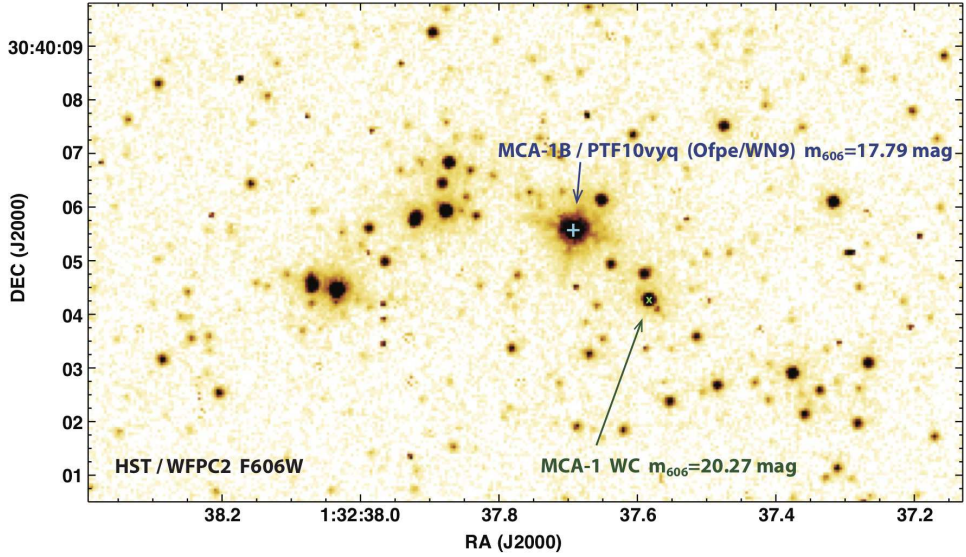}
\caption{An {\it HST}/WFPC2 image of the host cluster around MCA-1B in
  the F606W filter.  This image was taken by {\it HST} on 1994 July
  30 UT. The Ofpe/WN9 star MCA-1B (blue plus) and the WC star MCA-1
  (green $\times$) are identified.}
\label{fig:hst}
\end{figure*}

\begin{figure*}
  \includegraphics[width=5.6in]{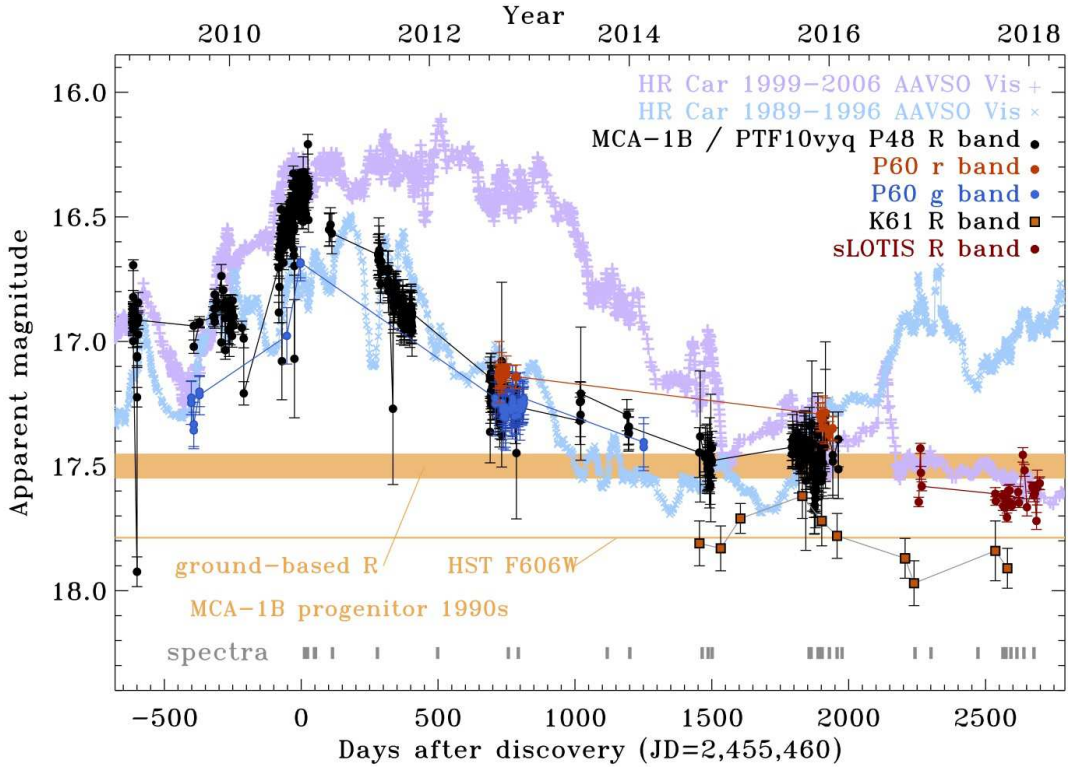}
\caption{The observed light curve of MCA-1B (PTF10vyq).  Apparent P48
  $R$-band magnitudes are plotted in black, P60 (Palomar 60-in)
  $g$-band magnitudes in blue, and P60 $r$ in red.  K61 $R$-band
  magnitudes are plotted in rust-coloured squares, and Super-LOTIS
  $R$-band photometry in burgundy circles. All of these measurements
  for MCA-1B/PTF10vyq are apparent magnitudes with no reddening
  correction.  For comparison, the orange horizontal bars denote the
  approximate $R$ magnitude of the progenitor star in the 1990s in
  ground-based images with good seeing (better than 1{\arcsec}) and
  the F606W magnitude observed by {\it HST} in 1994 (see text).
  Finally, the lavender and sky-blue light curves show visual
  photometry of the Galactic LBV star HR Carinae obtained from the
  AAVSO for comparison; the two colours correspond to two different
  ranges of dates in the same light curve.  These are shown as HR~Car
  might appear if it were in M33 with the same line-of-sight reddening
  as we have adopted for MCA-1B, scaling from the {\it Gaia} DR2
  distance \citep{smith19}.}
\label{fig:phot}
\end{figure*}

\begin{figure*}
  \includegraphics[width=5.6in]{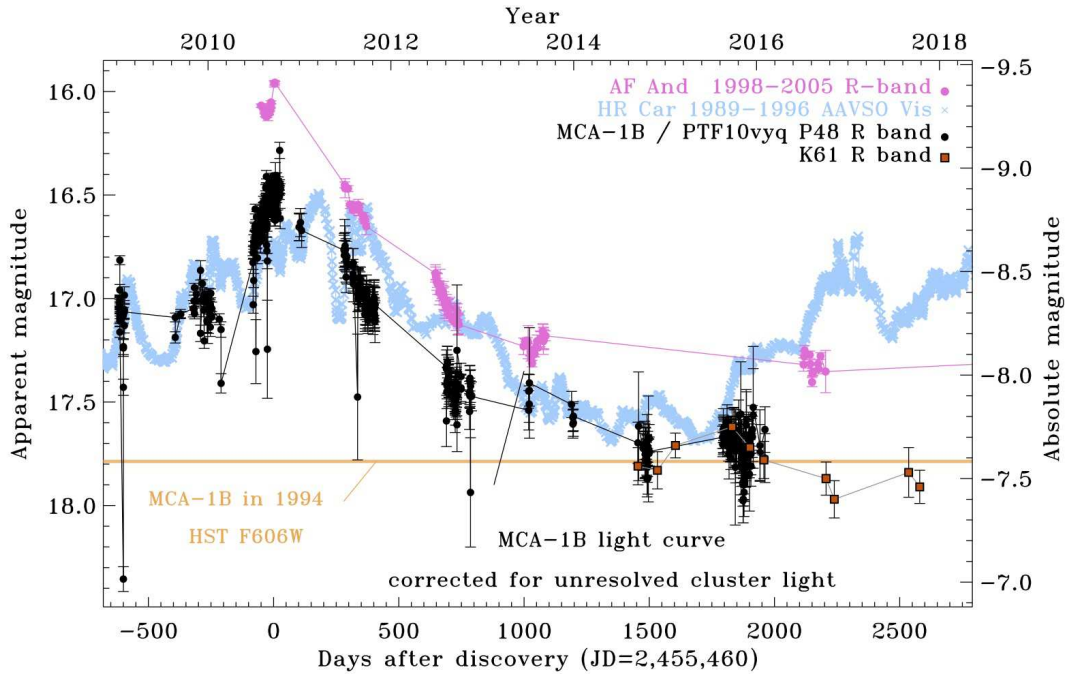}
  \caption{A simplified and adjusted version of the light curve in
    Figure~\ref{fig:phot}.  This uses the same PTF P48 light-curve
    data and the same K61 $R$-band data as in
    Figure~\ref{fig:phot}, except that we have subtracted a
    baseline amount of flux from P48 in an attempt to correct for
    contamination from unresolved neighbouring sources in the host
    cluster (see text and Figure~\ref{fig:img}). This is our best
    estimate of the true light curve corrected for contaminating
    light, guided by sources resolved in the {\it HST} image
    (Figure~\ref{fig:hst}).  Absolute magnitude is noted at right,
    assuming $D=830$ kpc and $E(B-V)=0.292$ mag.  The same AAVSO
    visual light curve of HR~Car over the interval 1989--1996 is shown 
    in blue for comparison.  Finally, the $R$-band light curve of the
    Hubble-Sandage variable AF~And (pink) is shown for comparison
    \citep{joshi19}, as if it were in M33 instead of M31 and with the
    same extinction as MCA-1B.}
\label{fig:phot2}
\end{figure*}

\begin{figure*}
\includegraphics[width=5.6in]{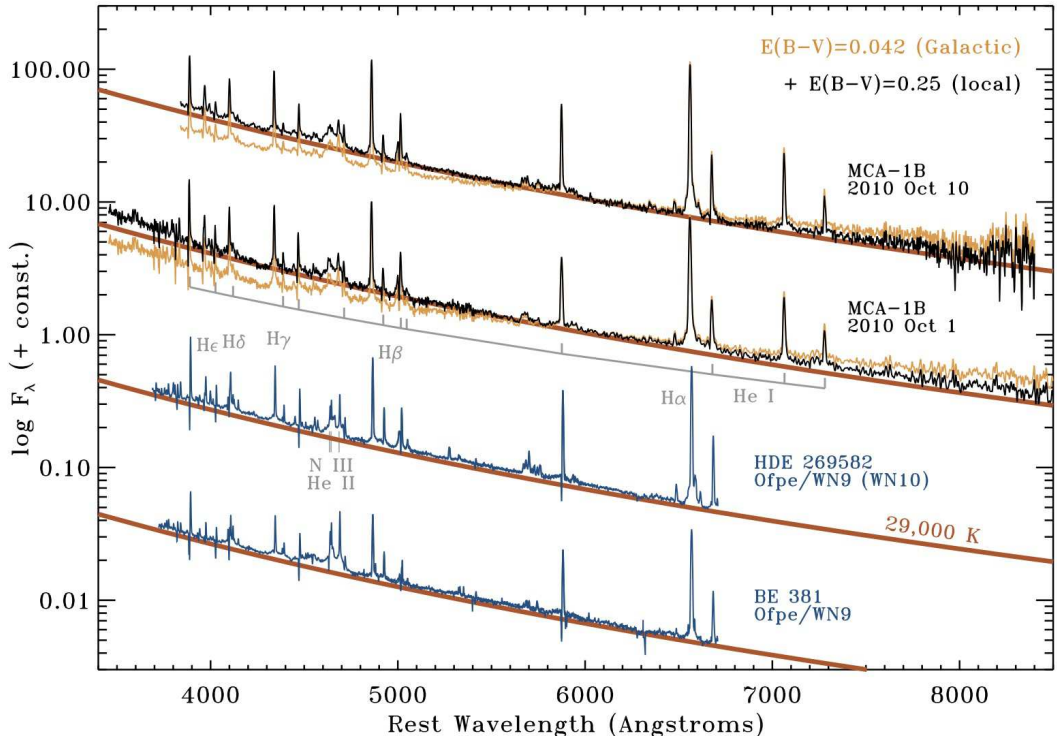}
\caption{Spectra of MCA-1B near maximum light shortly after
  discovery of the outburst.  The apparent spectral type is Ofpe/WN9
  (or WN9-11), and for comparison, the classic Ofpe/WN9 stars BE~381
  and HDE~269582 in the LMC are shown (blue).  These comparison
  spectra are from \citet{cs97}, with data kindly provided to the
  authors by P.\ Crowther.  The orange spectra of MCA-1B are the
  observed data corrected only for Milky Way line-of-sight reddening
  of $E(B-V)=0.042$ mag, and the black spectra have been corrected for
  a total reddening of $E(B-V)=0.292$ mag.  This is the amount needed
  to make the apparent continuum slope agree with the 29,000~K
  continuum temperature of a standard Ofpe/WN9 star with the same
  spectral type. With a Milky Way reddening in this direction of 0.042
  mag, this means the local reddening in this region of M33 is about
  0.25 mag.  We adopt a total reddening correction of 0.292 mag
  throughout this paper.}
\label{fig:ofpe}
\end{figure*}

\begin{figure*}
  \includegraphics[width=5.8in]{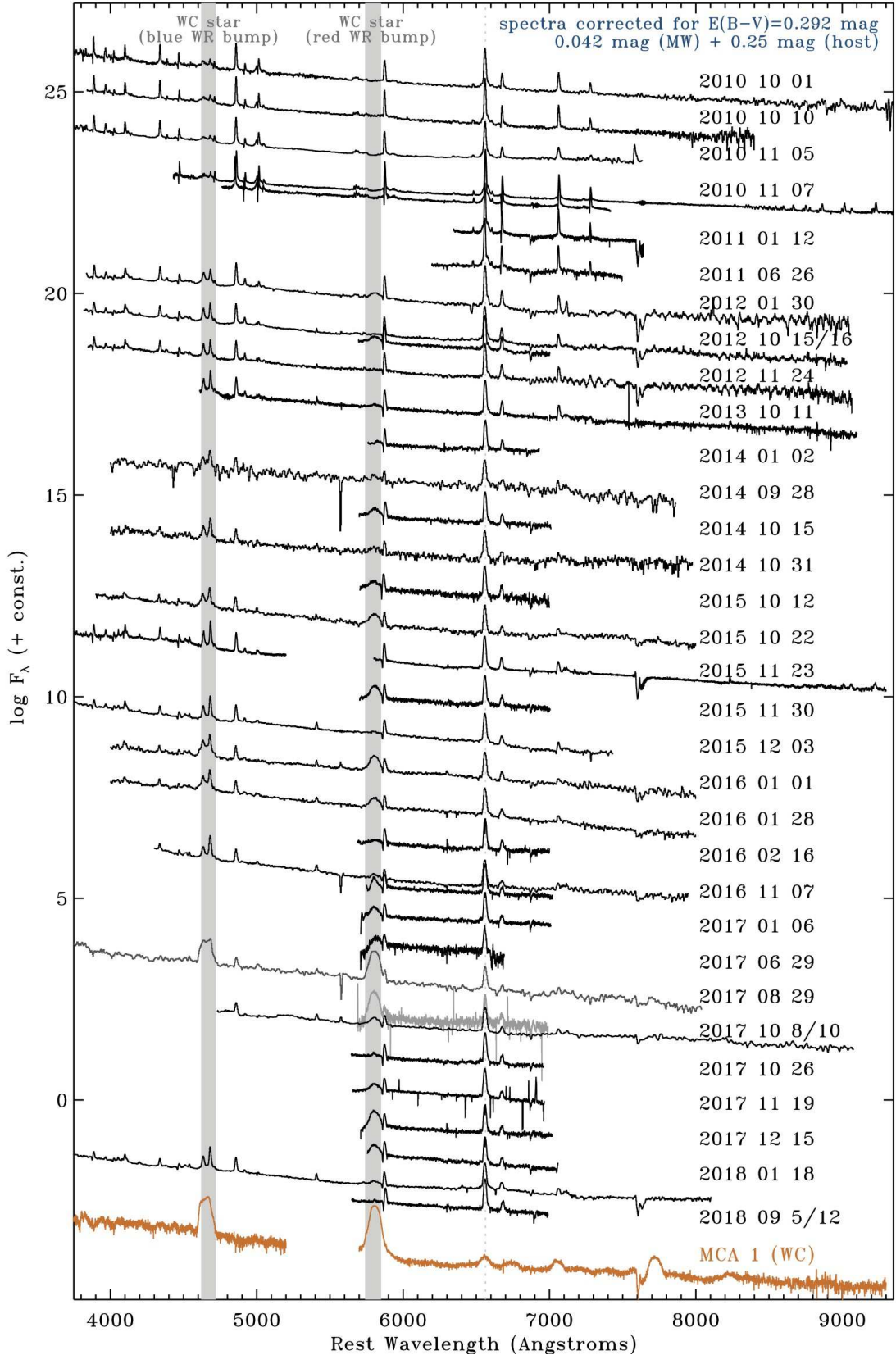}
\caption{Our full spectral sequence of MCA-1B from the time
  of discovery to 8~yr thereafter.  A total reddening
  correction of $E(B-V)=0.292$ mag has been applied to all spectra
  (see text). Grey bands denote the location of the blue and red WR
  bumps, for bright and broad emisison lines seen in WC stars.  A
  spectrum of the nearby WC star MCA-1 is shown at the bottom in
  orange.  The broad emission lines from this neighbouring star
  contaminate the spectrum of MCA-1B by differing amounts, depending
  on slit position, slit width, seeing, and the brightness of the
  primary target.}
\label{fig:spec}
\end{figure*}

\begin{table}\begin{center}\begin{minipage}{3.0in}
      \caption{Spectroscopy of MCA-1B (PTF10vyq) in M33}
\scriptsize
\begin{tabular}{@{}lcccc}\hline\hline
UT Date      &Day &Tel./Instr.  & grating & He~II$^a$ \\ 
   &   &   & (lines mm$^{-1}$) & \\ \hline
2010 Oct 01  &1       &Lick3m/Kast  &600/830 &* \\
2010 Oct 10  &10      &KPNO4m/RCSP  &316     &* \\
2010 Nov 05  &36      &Keck1/LRIS   &600/400 &* \\ 
2010 Nov 07  &38      &Keck2/DEIMOS &600     &* \\ 
2010 Nov 07  &38      &Keck2/DEIMOS &1200    & \\ 
2011 Jan 12  &104     &MMT/Blue     &1200    & \\
2011 Jun 26  &269     &MMT/Blue     &1200    & \\
2012 Jan 30  &487     &MMT/Blue     &1200    & \\
2012 Jan 30  &487     &MMT/Blue     &300     & \\
2012 Oct 15  &746     &MMT/Blue     &1200    & \\
2012 Oct 16  &747     &MMT/Blue     &300     & \\
2012 Nov 24  &786     &MMT/Blue     &300     &* \\
2013 Oct 11  &1107    &Keck2/DEIMOS &830        &* \\ 
2014 Jan 02  &1190    &MMT/Blue     &1200    & \\
2014 Sep 28  &1459    &Bok/BC       &300     &* \\
2014 Oct 15  &1476    &MMT/Blue     &1200    & \\
2014 Oct 31  &1492    &Bok/BC       &300     &* \\
2015 Oct 12  &1838    &MMT/Blue     &1200    & \\
2015 Oct 22  &1848    &Bok/BC       &300     & \\
2015 Nov 23  &1880    &LBT/MODS     &400/250 &* \\ 
2015 Nov 30  &1887    &MMT/Blue     &1200    & \\
2015 Dec 03  &1890    &MMT/Blue     &300     &* \\
2016 Jan 01  &1919    &Bok/BC       &300     & \\
2016 Jan 28  &1946    &Bok/BC       &300     & \\
2016 Feb 16  &1965    &MMT/Blue     &1200    & \\
2016 Nov 07  &2230    &Bok/BC       &300     &* \\
2016 Nov 07  &2230    &MMT/Blue     &1200    & \\
2017 Jan 06  &2290    &MMT/Blue     &1200    & \\
2017 Jun 29  &2464    &MMT/Blue     &1200    & \\
2017 Aug 29  &2556    &Bok/BC       &300     & \\
2017 Oct 08  &2565    &MMT/Blue     &1200    & \\
2017 Oct 10  &2567    &Bok/BC       &300     & \\
2017 Oct 26  &2583    &MMT/Blue     &1200    & \\
2017 Nov 19  &2607    &MMT/Blue     &1200    & \\
2017 Dec 15  &2633    &MMT/Blue     &1200    & \\
2018 Jan 18  &2667    &MMT/Blue     &1200    & \\
2018 Sep 05  &2900    &MMT/Blue     &300     &* \\ 
2018 Sep 12  &2907    &MMT/Blue     &1200    & \\ 
\hline
\end{tabular}\label{tab:spec}\end{minipage}
\end{center}
$^a$Observations with an asterisk in this column denote epochs with
  blue coverage when the seeing was good and resulting contamination
  from the neigboring WC star MCA-1 (judged by the broad red and blue
  WR bumps) was minimal.  These epochs were used to measure the
  strength of He~{\sc ii} $\lambda$4686 and other blue lines.
\end{table}

\section{DISCOVERY AND FOLLOW-UP OBSERVATIONS}

\subsection{PTF Discovery}

On 2010 Sep. 21 (UT dates are used throughout this paper; JD =
2,455,460.5), PTF10vyq was discovered in PTF images of M33 obtained in
the $R$-band filter with the Palomar 48-in telescope (P48) as part of
the PTF survey.  Even though a previously known source was present at
this position, the transient designation was triggered when it
brightened in difference images made by subtracting a template
constructed from images obtained in 2009.

This new transient source was named PTF10vyq, and was located in a
small star cluster at the centre of a faint shell-like H~{\sc ii}
region, located in the western outskirts of M33.  We discuss the
surrounding environment more below. A wide-field colour image from a
publically available Subaru/Suprime-Cam image showing its location in
M33 is included in Figure~\ref{fig:img}.  Figure~\ref{fig:hst} shows a
detail of the star cluster in which MCA-1B resides as seen in an
archival image taken with the {\it Hubble Space Telescope} ({\it HST})
Wide Field Planetary Camera 2 (WFPC2) in July 1994.

There were two previously known stellar sources within 2{\arcsec} of
each other that were near the position of the new transient PTF10vyq:
the WC star MCA-1 \citep{mca} and the Ofpe/WN9 star MCA-1B
\citep{willis92}.  These are unresolved in P48 images, and are only
marginally resolved in Kuiper 61-in (K61) images (see Figures~\ref{fig:img}c
and \ref{fig:img}d), but are clearly resolved by {\it HST}
(Figure~\ref{fig:hst}).  Since our first spectra (see below) revealed
that the transient resembled an Ofpe/WN9 star with narrow H emission
lines in the spectrum, we concluded that the transient source was
probably a brightening of the Ofpe/WN9 star MCA-1B, and was not
associated with an outburst of the WC star MCA-1.  Later images with
better seeing (where both sources were resolved) confirmed that this
was correct.  The Ofpe/WN9 star MCA-1B was already an LBV candidate
\citep{smith95}, and this new brightening provides another case of an
Ofpe/WN9 star that has transitioned into an eruptive state.

\subsection{1994 {\it HST}/WFPC2 Photometry}

The location of MCA-1B was observed in the F606W filter with {\it
  HST}/WFPC2 on 1994 July 30 (GO-5494; PI Bianchi).  We obtained this
image from the archive, and the reduced F606W image is shown in
Figure~\ref{fig:hst}.  From this image, we measure an apparent F606W
magnitude of $17.787 \pm 0.004$ mag for MCA-1B in July 1994 (Vega mag).
This is $\sim0.3$ mag fainter than ground-based $R$ magnitudes
estimated at a similar epoch ($R=17.5$ mag; \citealt{smith95}).  As
noted in Section~1, ground-based magnitudes were $V=17.7$ mag
in 1992 with 0$\farcs$8 seeing \citep{smith95} and $V=17.43$ mag in
1993 with 1$\farcs$4--1$\farcs$7 seeing \citep{massey96}.  The {\it
  HST} image does show several other fainter sources nearby; there are
a few within $\sim 1${\arcsec}, but there are several more stars (and
brighter ones) within $\sim 2${\arcsec}.  Although we cannot rule out
real photometric variability of MCA-1B over 1--2~yr timescales, these
nearby stars might contribute 10--20\% of the total light at that epoch
in a ground-based aperture, depending on the seeing (compare
Figure~\ref{fig:hst} to panels c and d of Figure~\ref{fig:img}).  For
example, the nearby WC Wolf-Rayet star MCA-1 has an apparent F606W
magnitude of $\sim20.3$ in this same {\it HST}/WFPC2 image. It is
fairly well resolved from MCA-1B in our K61 ground-based
images (Fig.~\ref{fig:img}d), but MCA-1 is blended with MCA-1B in PTF
P48 images (Fig.~\ref{fig:img}c) and in Super-LOTIS photometry (see
below).  The influence of these neighbouring stars is discussed in more
detail in Section~4.

\subsection{Follow-up Photometry}

We continued to monitor MCA-1B/PTF10vyq as part of the ongoing PTF
(and then iPTF) survey of M33 with the P48 in $R$ for the next
several years.  We also obtained some late-time images in the SDSS $g$
and $r$ bands with P60.  We obtained photometry from the PTF images
using point-spread-function (PSF) photometry with a 
3{\arcsec} PSF fitting radius.  As
mentioned above and described below, it is likely that the PTF
photometry has some contamination from several nearby point
sources in the host cluster.

We obtained late-time imaging photometry using the Mont4k camera on
K61 ($R$ band) starting in late 2014.  The
images were obtained in $3 \times 3$ binning mode, resulting in a final
scale of 0$\farcs$43 pixel$^{-1}$.  The K61 photometry is valuable in this
crowded region, since the typical image quality (1{\arcsec}) is better
than the median image quality of the P48, and the 0$\farcs$43 pixels
are smaller than the coarser 1{\arcsec} pixels of PTF.  Using
PSF-fitting photometry, we were able to clearly separate MCA-1B from
its neighbouring WC star MCA-1, as well as a few other, fainter point
sources (although not the faint neighbours within 1{\arcsec} in {\it
  HST} images).

We also obtained late-time $R$ photometry starting in
late 2016 using Super-LOTIS (updated Livermore Optical Transient
Imaging System), which is a 0.6-m robotic telescope on Kitt Peak that
uses a $2048 \times 2048$ CCD camera with 0$\farcs$5 pixels and a
17{\arcmin} field of view.\footnote{{\tt
    http://slotis.kpno.noao.edu/LOTIS/index.php}} Super-LOTIS images
have typical seeing of 2{\arcsec} and are comparable in quality to P48
images, so they also do not resolve MCA-1B from its immediate
neighbours in the host cluster.

A summary of the visual-wavelength photometry from PTF, K61, and
Super-LOTIS is provided in Figure~\ref{fig:phot}.  This figure also
denotes the apparent magnitude of the progenitor Ofpe/WN9 star in the
early/mid 1990s from both ground-based and {\it HST} photometry
(orange bars).  For comparison, we show the visible light curve of the
Galactic LBV star HR~Car during S~Doradus cycles in 1989--1996 (light
blue) and 1999--2006 (lavender). These are apparent visual magnitude
estimates of HR~Car from the American Association of Variable Star
Observers (AAVSO).\footnote{\tt https://www.aavso.org/lcg}

Since MCA-1B is in a crowded region, the apparent magnitude is
contaminated by neighbouring stars in ground-based photometry, with
different amounts of contamination for different image quality and
pixel scales.  Figure~\ref{fig:phot2} shows a simplified version of
the light curve of MCA-1B, where we have attempted to correct for
contamination by subtracting a baseline flux estimated from neigbouring
stars in the {\it HST} F606W image.  This is discussed further
below. While the precise amount of contamination is difficult to
quantify, Figure~\ref{fig:phot2} gives a better representation of the
true visible-light variations of MCA-1B's outburst than does
Figure~\ref{fig:phot}.

\subsection{Follow-up Spectroscopy}

A few weeks after PTF's discovery of the September 2010
brightening, we obtained visual-wavelength spectra of MCA-1B/PTF10vyq
using the Kast spectrograph \citep{ms93} on the Lick 3~m Shane
reflector (2010 Oct. 1), and using the RC Spectrograph\footnote{{\tt
    https://www.noao.edu/kpno/manuals/rcspec/rcsp.html}} on the Kitt
Peak National Observatory (KPNO) 4~m Mayall telescope (2010 Oct. 10).
These two spectra were obtained near the time of peak brightness, and
both covered the full optical range from below 4000 \AA\ to about 1
$\mu$m with moderate spectral resolution ($R \approx 700$).  They were
reduced using standard tasks in IRAF, and the resulting calibrated
spectra are shown in Figure~\ref{fig:ofpe}.  

Subsequently, a long
series of follow-up spectra spanning several years was obtained using
the Boller \& Chivens (B\&C) Spectrograph mounted on the 2.3~m Bok
telescope on Kitt Peak, and the Bluechannel (BC) spectrograph on the
6.5~m Multiple Mirror Telescope (MMT).  The Bok spectra were all
obtained with a low-resolution grating (300 lines mm$^{-1}$; $R \approx 700$)
covering most of the optical range, whereas the MMT spectra
used either low resolution (300 lines mm$^{-1}$ grating; $R \approx 500$) over a
wide wavelength range (4000--8000 \AA) or moderately high resolution
(1200 lines mm$^{-1}$ grating; $R \approx 4500$) over a smaller range (typically
5700--7100 \AA, although at some epochs a different central wavelength
was chosen).  We also obtained spectra using the Low Resolution
Imaging Spectrometer (LRIS; \citealt{oke95}) and the Deep Imaging
Multi-Object Spectrograph (DEIMOS; \citealt{faber03}) at Keck
Observatory.  Finally, we obtained one late spectral epoch (2015
Nov. 23) using the Multi-Object Double Spectrograph (MODS;
\citealt{bb00}) on the Large Binocular Telescope (LBT). All spectra
were obtained with the long slit at the parallactic angle
\citep{filippenko82}, to minimise the effects of atmospheric
dispersion.  Details of
the spectral observations are listed in Table~\ref{tab:spec}, and
the spectra are plotted in Figure~\ref{fig:spec}.

Figure~\ref{fig:ofpe} shows our two spectra of MCA-1B shortly after
the discovery of PTF10vyq.  Although there may be some minimal
contamination from the red and blue Wolf-Rayet (WR) bumps, the spectra
at times near peak brightness are clearly consistent with standard
Ofpe/WN9 stars (BE~381 and HDE~269582 in the LMC are shown for
comparison).  This hot spectrum at peak brightness is very unusual for
LBVs.  These spectra of MCA-1B are shown corrected for Milky Way
line-of-sight reddening of $E(B-V)=0.042$ mag (orange) and also for an
additional amount of local host reddening (black). The value shown is
an extra reddening correction of $E(B-V) = 0.25$ mag (for a total of
$E(B-V) = 0.292$ mag).  This is the amount chosen to make the
continuum of PTF10vyq match the continuum slope of a typical Ofpe/WN9
star at 29,000~K.  Values of 28--30~kK are usually quoted for Ofpe/WN9
stars.
\footnote{The hot temperature at peak brightness is inferred from the
  Ofpe/WN9-like spectral properties at that time (including strong
  He~{\sc i} emission lines), not from the shape of the dereddened
  spectrum.  The choice of $E(B-V)$ therefore does not influence the
  main result that MCA-1B stays hot at peak.}  The continuum slope and
$T_{\rm eff}$ may differ slightly due to mass-loss rate and local
reddening of course; the precise value of $T_{\rm eff}$ here is less
important than the fact that it showed little change.

Figure~\ref{fig:spec} shows the full series of optical spectra we
obtained of MCA-1B, all corrected for $E(B-V)=0.292$ mag of reddening.
Line profiles of the strong, narrow emission lines H$\alpha$, He~{\sc
  i} $\lambda$6678, and He~{\sc i} $\lambda$5876 are shown in
Figures~\ref{fig:ha}, \ref{fig:he6680}, and \ref{fig:he5876},
respectively. In these figures, we favour epochs with higher spectral
resolution and we omit redundant epochs of lower resolution spectra.
Many of the spectra are contaminated by light from the nearby WC star
MCA-1, especially at later times.  Although the WC star is much
fainter in the continuum than MCA-1B, its broad emission lines are
bright, and some of this emission was unavoidably included in
extractions of the spectrum of our main target.  The amount of
contamination from the WC star varies depending on the exact slit
position angle and seeing. Although we always set the slit at the
parallactic angle, the resulting position angle on the sky varied.
The relative contamination is barely noticable at early times when the
transient was bright, but it worsened as MCA-1B faded.

For reference, we include an LBT/MODS spectrum of the WC star
taken on 2015~Nov.~23 (shown in orange in Figure~\ref{fig:spec}); this
was taken on a night with good seeing when tracings of the two sources
could be reliably separated, and when we placed the slit at a position
angle that included both targets.  The brightest contaminating
features are the blue and red WR emission bumps in WC stars, which are
marked with grey vertical bands to help identify this contamination in
all the spectra of MCA-1B.  Note that the WC star MCA-1 does {\it not}
exhibit bright, broad emission lines of He~{\sc i} or H$\alpha$, so
contamination from MCA-1B does not contribute to the broadening of
those emission lines as MCA-1B faded, which we report in this
paper.  Epochs judged to have minimal contamination from the nearby WC
star are denoted with an asterisk in the last column of Table 1.

\section{LOW-METALLICITY ENVIRONMENT}

MCA-1B resides in the western outskirts of M33
(Figure~\ref{fig:img}a), where the metallicity gradient implies a
relatively low local metallicity.  MCA-1B is located about 5.8--6 kpc
from the centre of M33, correcting for the projected inclination.  It
is close (a few arcmin on the sky) to the H~{\sc ii} region LGC~HII~3,
for which \citet{magrini07} report an oxygen abundance of
12 + log(O/H) = $8.24 \pm 0.05$, consistent with the value of
12 + log(O/H) = 8.2 expected from their fit to the overall metallicity
gradient at a radius of $\sim 6$ kpc.  This translates to about
0.35~$Z_{\odot}$, based on an adopted solar oxygen abundance of
12 + log(O/H) = 8.69 \citep{asplund}, placing the local metallicity of
MCA-1B in between that of the LMC and SMC.  This relatively low
metallicity might be relevant for interpreting peculiarities of its
eruption.

Although MCA-1B is in the somewhat remote western regions of M33, it
is not completely isolated; it is found amid a cluster or association
of a dozen or so luminous blue stars.  The cluster is roughly at the
centre of a large and evolved H~{\sc ii} region shell nebula
(Figure~\ref{fig:img}b).  This shell has a diameter of roughly
60{\arcsec}, or about 240~pc.  The shell-like H~{\sc ii} region is
faint and filamentary, and relatively devoid of H$\alpha$ emission in
its interior (see Figure~\ref{fig:img}b); this implies that it is an
old and evolved H~{\sc ii} region.  With a radius of 120 pc and
expanding with a typical speed of 15--20 km s$^{-1}$, this shell would
have an age of (very roughly) 6--8~Myr, commensurate
with stellar lifetimes for initial masses of 20--30
$M_{\odot}$. 

It is interesting that MCA-1B is in a relatively young ($<10$~Myr) and
blue cluster, because this is unlike most LBVs \citep{st15}.  Given
its high luminosity, however, it may be overluminous for the age of
its host cluster, depending on the true age of the region; the rough
age estimate of the nebula noted here is not precise enough to
determine if it is a massive blue straggler.  Future imaging
photometry of the cluster with {\it HST} may allow a more precise
isochrone age, which could reveal if MCA-1B is a massive blue
straggler like other LBVs.  An initial mass of 50 $M_{\odot}$ was
implied by the luminosity of the quiescent progenitor star as compared
to single-star evolutionary tracks \citep{smith95}.

It is useful to place the environment of MCA-1B in context.  Images
reveal a surrounding H~{\sc ii} region shell and an associated star
cluster that is crowded enough to corrupt the ground-based photometry,
as discussed next.  These clues about a crowded environment, however,
follow from the luxury that M33 is less than 1~Mpc away.  If MCA-1B
were in a more distant galaxy that is typical of supernova (SN) 
studies, we might cast its surroundings somewhat differently.

Take the case of SN~2009ip, for example, which has been noted to reside
in a remarkably isolated environment \citep{smith+16}.  This is
surprising for its presumably high-mass, LBV-like progenitor.
SN~2009ip is found about 5~kpc from the centre of its modest spiral
host galaxy, similar to the location of MCA-1B at $\sim 6$ kpc from
the centre of M33.  Deep late-time {\it HST} images did not resolve a star
cluster and did not detect a bright extended H~{\sc ii} region around
SN~2009ip \citep{smith+16}.  However, the host galaxy of SN~2009ip,
NGC~7259, is at a much larger distance of 20.4~Mpc, about 25 times
farther away than M33.  If MCA-1B and its environment were moved to
that distance, it is doubful that an {\it HST} image would detect the
surrounding faint stars if MCA-1B exploded --- the whole cluster would
only be a few {\it HST} pixels across, and could easily be hidden in
the glare of the fading SN.  At 25 times more distant, the thin H~{\sc
  ii} region shell around MCA-1B might easily fall below the surface
brightness detection limits in {\it HST} images like the ones used to
search for H~{\sc ii} regions around SN~2009ip, which could have
detected large and bright nebulae like the Carina Nebula or 30~Dor,
but not much fainter ones.  The associated cluster and H~{\sc ii}
regions would certainly not be detectable in typical ground-based
images if MCA-1B were located so far away.  Thus, despite the issues
with a crowded environment noted below, it is worth remembering that
MCA-1B is nevertheless characteristic of some of the more remote
environments found for core-collapse supernovae (SNe).

\section{IN A CROWDED REGION}

MCA-1B resides in a small cluster or association with several
neighbouring stars within a few arcseconds (Figure~\ref{fig:hst}).
This means that its nearest neighbours potentially impact our
photometric and spectroscopic observations, depending on the relative
brightness as MCA-1B changes, but also depending on seeing, pixel
scale, and photometry parameters, as well as spectroscopic slit width
and orientation.  This complicates the interpretation of the light
curve and spectral evolution.  The crowded region is also interesting
from the perspective that it is actually quite unusual for LBVs; most
LBVs appear to shy away from crowded regions with clusters of O-type
stars \citep{st15}.  This is thought to be the result of binary
evolution, either through kicks from a companion's SN explosion or by
rejuventation\footnote{``Rejuvenation''
  \citep{dt07,demink13,justham14,schneider16} refers to a star that
  accretes through mass transfer in a binary or undergoes a merger,
  evolving thereafter as if it were a younger, more massive, rapidly
  rotating, and more luminous star (i.e., a massive blue straggler). }
\citep{mojgan17,st15,smith16,smith19bbs}.  Interestingly, \citet{st15}
also found that WC stars were more isolated than expected from
single-star evolution.  Here we have both an LBV and a WC star in the
same cluster, offering the opportunity to constrain the age of the
cluster and thus the initial mass of both the LBV and WC star.

\subsection{Crowded photometry}

The most direct impact of this contaminating light is on the
photometry and light curve.  For PSF-fitting photometry in the PTF
images, which had 2{\arcsec} typical image quality and 1{\arcsec}
pixels, and used an initial 3{\arcsec} PSF-fitting radius, there is
more contaminating flux from neighbouring stars than in the K61
images with 1{\arcsec} typical image quality and 0$\farcs$43 pixels
(compare Figures~\ref{fig:img}c and \ref{fig:img}d).  This is why the
late-time PTF magnitudes are brighter than K61 magnitudes in
Figure~\ref{fig:phot}.

We are interested in the true variability amplitude of MCA-1B, so we
must make some correction for contaminating light.  The sharpest
available image is the {\it HST}/WFPC2 F606W image taken in 1994
(Figure~\ref{fig:hst}).  It is clear that there are several
neighbouring stars within 2--3{\arcsec}.  We experimented with
photometry in various-sized apertures and with PSF fitting on blurred
images to make a rough assessment of the contamination that might be
present in the different ground-based images.  Depending on
conditions, we find it likely that roughly 30\% of the total flux in
PTF and super-LOTIS images at late times is contributed by
neighbouring stars (requiring a correction of $+$0.3 mag).

Guided by this likely contamining flux estimated from the {\it HST}
image, we therefore adopted a baseline flux level to subtract from all
of the PTF photometry.  This is not a precise correction because it is
a different filter, and because the seeing and therefore contamination
varied from one epoch to the next.  However, the variability timescale
is much slower than our typical PTF cadence, so the average trend in
the data gives a good representation of the average light curve.  This
chosen correction allows the K61 and PTF photometry to agree at late
times.  The resulting adjusted light curve from the PTF and K61
photometry is shown in Figure~\ref{fig:phot2}.  We think that the
light curve in Figure~\ref{fig:phot2} is a much better representation
of the true variability of MCA-1B than the raw photometry with no
correction in Figure~\ref{fig:phot}.

There are two key consequences of including these corrections.  First,
the full amplitude of variability is greater, because the relative
correction is small when MCA-1B is bright and larger when it has
faded.  Second, with this correction applied, it appears that in the
last few years, MCA-1B has settled down to be roughly comparable to
its pre-outburst quiescent magnitude (or even fainter), signaling the
end of its current eruption.  Compared to its state in the early 1990s
and its post-outburst quiescent state, MCA-1B brightened by about 1.4
mag.  The amplitude is comparable to that of classical LBV eruptions
of S~Doradus variables.  We show the light curve of the classic
Galactic LBV star HR~Car again for comparison in
Figure~\ref{fig:phot2}, although there is considerable diversity among
the light curves of LBVs (see, e.g., \citealt{vangenderen01}).  We
also show the $R$-band light curve of the classic Hubble-Sandage
variable AF~And in M31 \citep{joshi19}.  AF~And in M31 and MCA-1B in
M33 are remarkably similar in terms of their peak amplitude, duration,
decay rate, and long post-eruption quiescence.

\subsection{Crowded spectroscopy}

The other main observational effect of the crowded region is its
impact on the spectrum, which changes depending on observing
conditions and instrument parameters.  Nearby main-sequence OB stars
in the host cluster don't matter very much for the appearance of the
spectrum; they are relatively faint at visible wavelengths and have
mostly hot continuum plus some weak absorption lines.  This
contamination can influence the measured line equivalent widths (EWs),
just as it does the broadband photometry, because it adds continuum
flux without adding emission-line flux.  We found significant
night-to-night variations in line EW measurements, and it was
generally true that the EWs were relatively weaker (up to about 40\%
in some cases) on nights with poorer seeing; this is as expected,
because there would be more contaminating continuum light from the
cluster when the seeing is bad.  The LBV spectrum is mostly free from
contamination at early times when it outshines the cluster light.

In terms of spectral features, the most important source of
contaminating light that influenced the appearance of the spectrum is
MCA1, a luminous WC star located 2{\arcsec} away.  This star has very
strong, broad emission lines that can be seen in two-dimensional
long-slit spectra when the slit position angle includes light from
this star, or when the seeing was relatively poor, as noted earlier.

In general, the relative contamination of the spectrum by the WC star
grows with time as the transient fades slowly, although it varies
considerably from one spectroscopic observation to the next.  The most
noticable contaminating features in the spectrum are the blue and red
``WR bumps'' that are known to be strong emission features in WC
stars.  These are due primarily to He~{\sc ii} $\lambda$4686 in the
blue WR bump and C~{\sc iv} $\lambda$5808 in the red WR bump.  These
are noted by grey vertical shaded bars in Figure~\ref{fig:spec}.  We
also show the spectrum of the neighbouring WC star itself at the
bottom of Figure~\ref{fig:spec} in orange.  The continuum slope of the
WC star is roughly the same as that of the primary target (both are
sampling the Rayleigh-Jeans tail of a hot star), so its contaminating
continuum light does not alter the apparent shape of the continuum for
the LBV.  Looking through the series of spectra of MCA-1B in
Figure~\ref{fig:spec}, one can see that some epochs have strong
contamination from the WR bumps, especially on 2016 Jan. 1, 2017
Aug. 29, 2017 Oct. 8/10, and 2017 Dec. 15, for example.  Overall,
then, the apparent changes in the broad WR bumps at late times should
be ignored by the reader.  The red WR bump is especially noticable,
and is present to some degree in all the spectra after the early-time
peak.  In the early epochs during the main peak in 2010, the
contamination from the neighbouring WR star is negligible, because the
LBV is much brighter.

Narrow emission lines in the spectrum are all due to the LBV, and
their profiles are largely unaffected by the contaminating cluster or
WC star light in the spectrum.  (Their equivalent widths, however,
could be affected if there is extra continuum included in the slit; we
return to this later in Section 5.2.)  No narrow lines are seen in the
WC star on nights with good seeing when its spectrum can be well
separated, and no broad lines are seen in the LBV on those same nights
except for moderately broad electron-scattering wings of strong narrow
lines.

The strong WC contamination of the red WR bump, however, complicates
the measurement of the narrow He~{\sc i} $\lambda$5876 line that we
discuss below. This is exacerbated by the consistent presence of P
Cygni absorption in He~{\sc i} $\lambda$5876.  These effects can alter
the measured EW and full width at half-maximum intensity (FWHM) of the
narrow He~{\sc i} $\lambda$5876 emission.  We note appropriate caution
below.

\begin{figure}
  \includegraphics[width=2.7in]{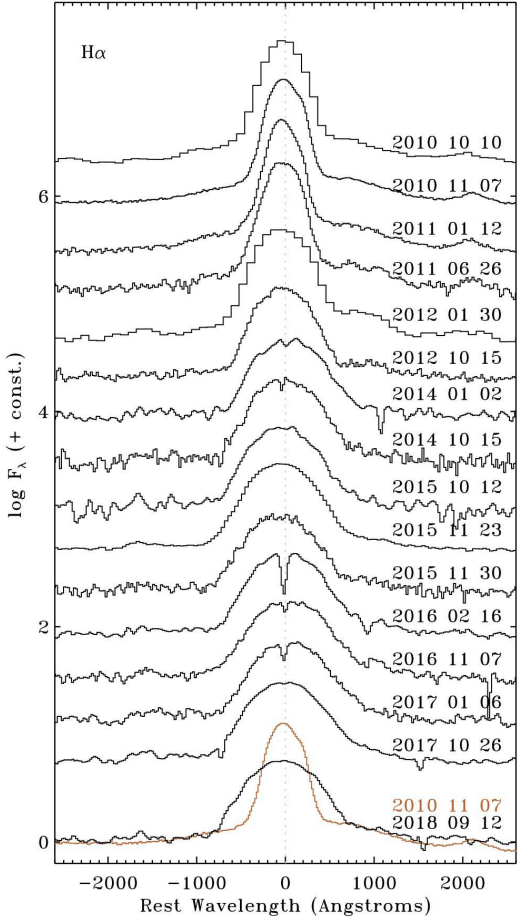}
\caption{A detail of the H$\alpha$ line profile in MCA-1B at various
  epochs.  We have chosen to exclude some of the lower resolution or
  lower signal-to-noise ratio spectra, since the changes tend to be
  subtle and slow over time. Line profiles are shifted and aligned by
  their flux centroids to emphasise changes in line profile shape and
  width.  An early-epoch spectrum with a narrower line profile is
  reproduced in orange at the bottom for direct comparison with the
  late-time profile.}
\label{fig:ha}
\end{figure}

\begin{figure}
  \includegraphics[width=2.7in]{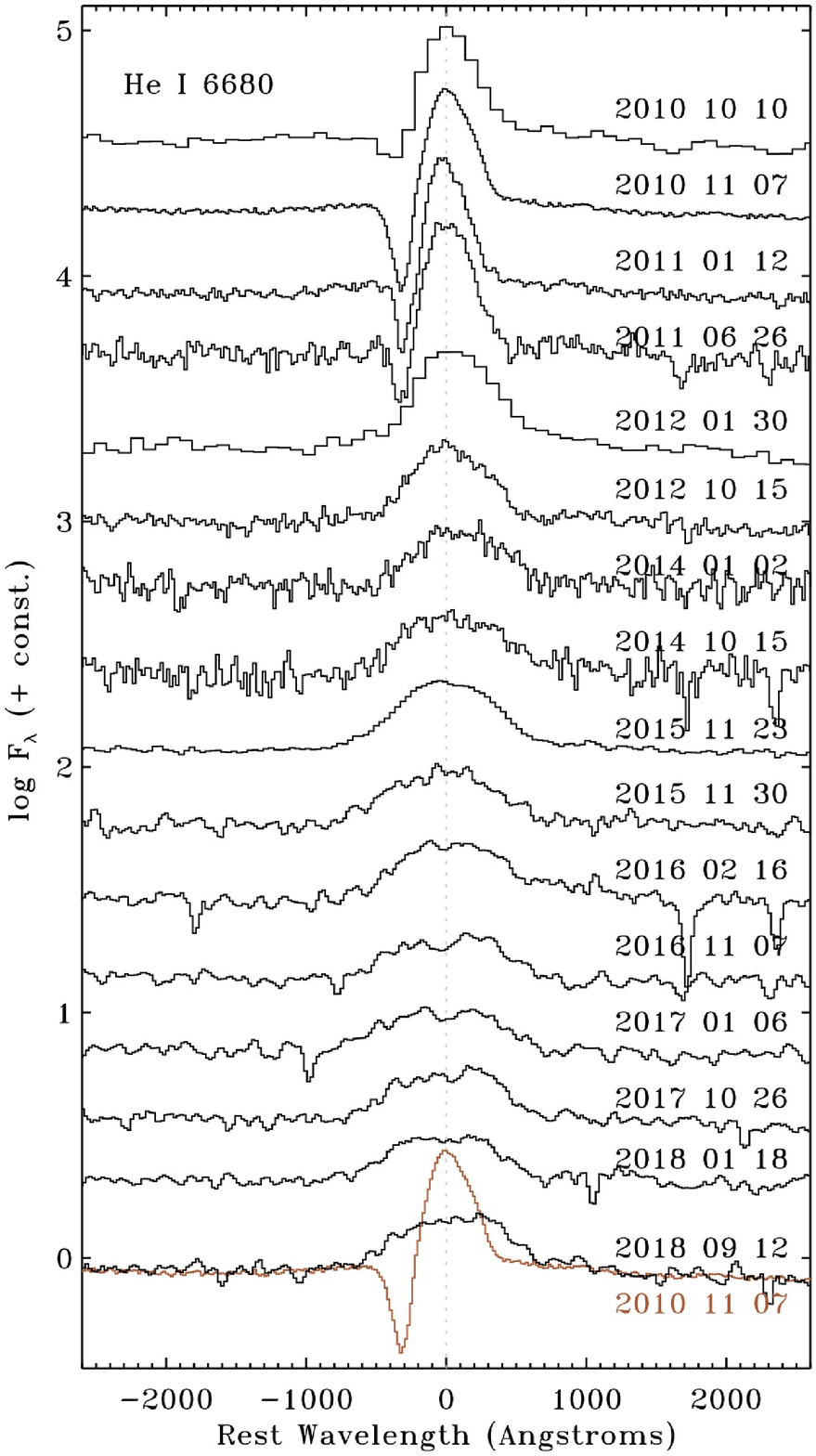}
\caption{Same as Figure~\ref{fig:ha}, but for the evolution of the
  He~{\sc i} $\lambda$6678 line profile.}
\label{fig:he6680}
\end{figure}

\begin{figure}
  \includegraphics[width=2.7in]{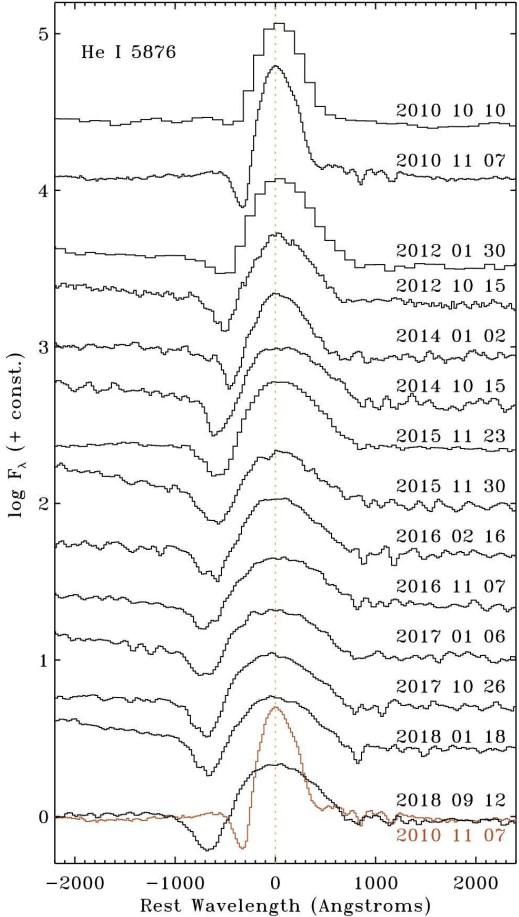}
\caption{Same as Figure~\ref{fig:ha}, but for the evolution of the
  He~{\sc i} $\lambda$5876 line profile.}
\label{fig:he5876}
\end{figure}

\section{Physical Parameters}

\subsection{The 1990s progenitor}

Prior to the PTF discovery of eruptive variability in 2010, MCA-1B was
not known to be significantly variable.  The quiescent progenitor star
was, however, subjected to close scrutiny and quantitative
spectroscopic analysis.

Based on detailed studies and quantitative analysis of the spectrum in
the 1990s, physical parameters have been estimated for the quiescent
Ofpe/WN9 progenitor star MCA-1B.  \citet{crowther95} estimated
log($L/L_{\odot}$) = 5.9 and $T_{\rm eff}$ = 29~kK, and in a related study,
\citet{smith95} similarly estimated log($L/L_{\odot}$) = 5.84, $T_{\rm
  eff}$ = 29~kK, $\dot{M} = 10^{-4}$~$M_{\odot}$ yr$^{-1}$, and a
terminal wind speed of $v_{\infty}$ = 420 km s$^{-1}$.  In both studies,
the derived bolometric luminosity depends on the effective
temperature, but is tied to the visual magnitude.  Both of these
studies adopted $V = 17.7$ mag and $R = 17.5$ mag.

However, as noted above, a comparison of {\it HST} and various
ground-based images reveals some contamination from nearby stars in
ground-based photometry.  Adopting instead an approximate $R$-band
apparent magnitude of 17.79 indicated by {\it HST} imaging around the
same time, the luminosity would be scaled down by $\sim 0.3$
mag or $\sim 0.12$ dex, to about log($L/L_{\odot}$) = 5.72.  Based on
comparing the luminosity to single-star evolution models,
\citet{smith95} estimated that this corresponds to an initial mass for
MCA-1B of around 50 $M_{\odot}$, although with the lower luminosity
that has been corrected for contaminating light of nearby stars, the
initial mass implied by single-star evolutionary models would be more
like 40--45 $M_{\odot}$.

\begin{figure}
  \includegraphics[width=2.9in]{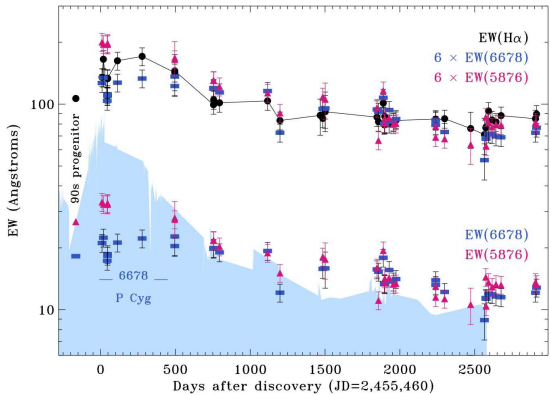}
\caption{MCA-1B's emission-line EW (with positive
  values for emission), measured for H$\alpha$ (black dots and lines),
  He~{\sc i} $\lambda$6678 (blue rectangles), and He~{\sc i}
  $\lambda$5876 (magenta triangles).  For close comparison with
  H$\alpha$, the He~{\sc i} lines are also shown multiplied by a
  constant value of 6.  The light-blue shaded region in the background
  is a representation of the light curve on the same timescale for
  reference.}
\label{fig:ew}
\end{figure}

\subsection{The hot temperature and bolometric luminosity at peak}

The overall spectral evolution of MCA-1B is shown in
Figure~\ref{fig:spec}, while details of the line-profile variation of
H$\alpha$, He~{\sc i} $\lambda$6678, and He~{\sc i} $\lambda$5876 are
shown in Figures~\ref{fig:ha}, \ref{fig:he6680}, and \ref{fig:he5876},
respectively.  The temporal evolution of emission-line
EWs of He~{\sc i} $\lambda$5876, $\lambda$6678, and H$\alpha$ is
shown in Figure~\ref{fig:ew}.

The spectral evolution of MCA-1B exhibits subtle changes in the
intrinsic spectrum, at least compared to the much more dramatic
temperature shifts usually seen in LBVs.  It retains a late WN-type
spectrum throughout its eruption.  The continuum slope remains
constant throughout the eruption, from which we infer that the fading
from peak is not attributable to dust formation.  There are some
apparent changes in Figure~\ref{fig:spec} that can, however, be
attributed to contamination from nearby sources if we account for two
effects that have been noted above.

First, the emission-line morphology is dominated by the narrow-line
spectrum of the LBV, except for the two WR bumps. As noted earlier,
these broad features are due to contamination from the nearby WC star
MCA-1.  This is clearly evident when one considers the fact that on
certain dates with good seeing conditions, such as the LBT spectrum on
2015 Nov. 23, these broad WR bumps are absent from the LBV spectrum,
and on that same date it is clear that those lines are fully
attributable to the WC star.  Ignoring contamination from these broad
WC lines, the morphology of the spectrum remains consistent with a hot
Ofpe/WN9 star at peak, retaining a late-WN type at all times.

Second, we know from comparing {\it HST} and ground-based imaging
photometry, discussed above, that as the LBV fades, there is an
increase in the relative contamination of {\it continuum} light from
unresolved neighbours.  This contamination is negligible at peak when
the LBV is bright, but we estimated that in ground-based data,
unresolved neighbours may contribute roughly 30\% of the
visible-wavelength continuum at late times.  This extra continuum can
cause a quantitative change in the measured EWs with time, but will
vary with seeing conditions and slit widths of the spectroscopic
observations.  Even if the intrinsic spectrum of the LBV had no change
from peak to quiescence, we would expect that the emission-line EWs of
various lines should decrease at late times.  This is in fact what is
observed, except that the change is larger (about a factor of 2) than
can be explained by 30\% contaminating continuum light alone.
Examining the time dependence of EWs for He~{\sc i} $\lambda$5876,
$\lambda$6678, and H$\alpha$ (Figure~\ref{fig:ew}), the measured EWs
decrease in-step by about a factor of 2 from peak to quiescence.
Emission lines are indeed stronger in brighter phases of the eruption.

The ratio of H$\alpha$ to these He~{\sc i} lines is roughly constant
as the LBV fades.  In addition to measured EWs, Figure~\ref{fig:ew}
shows EWs of the two He~{\sc i} lines scaled up by a factor of 6.
This artificial shift makes their EWs appear comparable to that of
H$\alpha$ for comparison.  As the LBV fades, this relative offset
between He~{\sc i} lines and H$\alpha$ stays roughly the same.

\begin{figure}
  \includegraphics[width=3.2in]{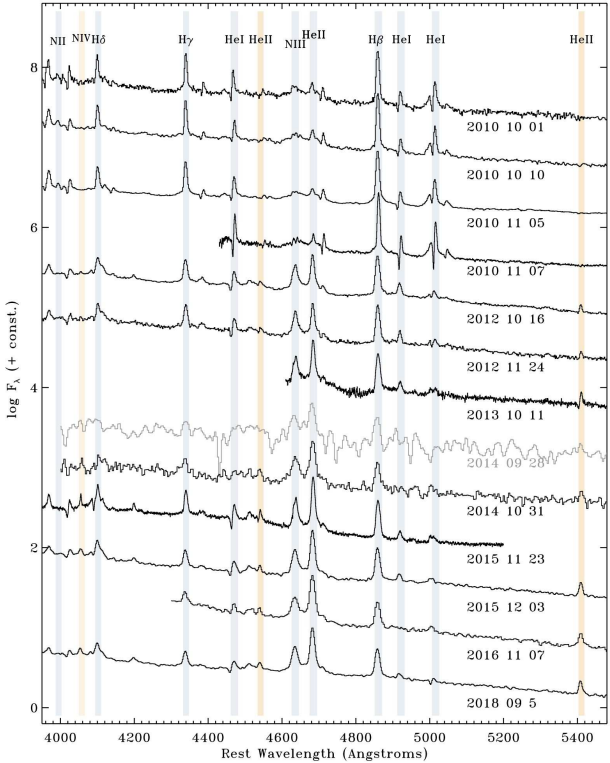}
\caption{Similar to Figure~\ref{fig:spec}, but concentrating on the
  blue part of the spectrum, and showing only spectra without strong
  contamination by the nearby WC star (epochs marked by ``*'' in Table
  1).  Several lines are marked by shaded vertical bars.  Note the
  change in relative strength of He~{\sc ii} and N~{\sc iv} lines
  (weak at peak, stronger at late times) as compared to Balmer lines
  and He~{\sc i} lines (weaker at late times).  In particular, the
  presence of emission from He~{\sc ii} $\lambda$4542 and
  $\lambda$5411 as well as N~{\sc iv} $\lambda$3995 at late times
  (shown with light-orange vertical bars) suggest a transition to
  a hotter spectral type of WN7--WN8.}
\label{fig:heii}
\end{figure}

\begin{figure}
  \includegraphics[width=2.9in]{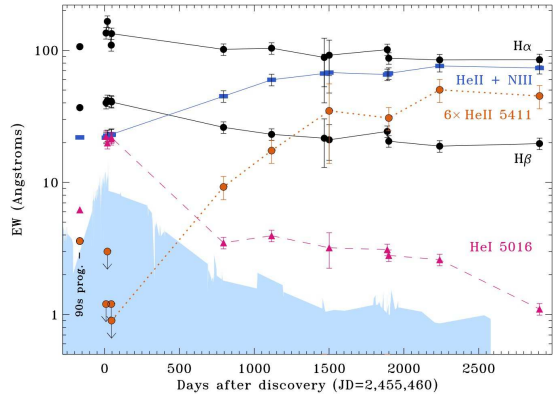}
\caption{Same as Figure~\ref{fig:ew}, but showing emission-line
  equivalent widths for lines in the blue spectrum.  These are
  measured on a subset of spectral epochs with blue wavelength
  coverage, when the seeing was sufficiently good that contamination
  from the neighbouring WC star was judged to be minimal (see Table
  1).  H$\alpha$ is the same as in Figure~\ref{fig:ew}.  Other lines
  are H$\beta$, the full blend of He~{\sc ii} $\lambda$4686 + N~{\sc
    iii} $\lambda$3434 (i.e., the blue WR bump), He~{\sc i}
  $\lambda$5016, and He~{\sc ii} $\lambda$5411.  He~{\sc ii}
  $\lambda$5411 is a weak line and is therefore multiplied by a factor
  of 6 for display, and its first few epochs are upper limits.  Values
  reported for the progenitor in the 1990s \citep{smith95} are also
  shown.}
\label{fig:ewhe2}
\end{figure}

\begin{figure}
  \includegraphics[width=2.9in]{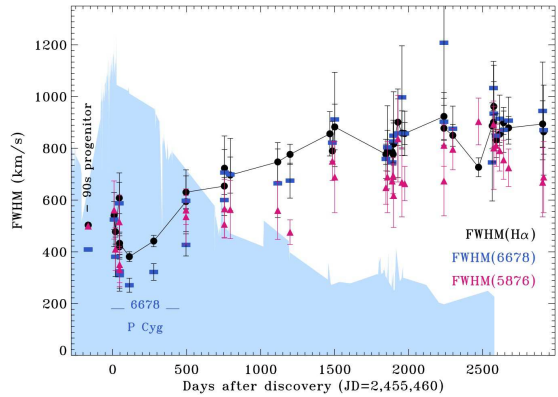}
\caption{Same as Figure~\ref{fig:ew}, but for the FWHM
  velocity of each line, derived from a Gaussian fit to
  the narrow emission component (ignoring the broad electron-scattering 
  wings).  The He~{\sc i} $\lambda$5876 emission FWHM is
  consistently lower because of its persistent P Cygni absorption.
  He~{\sc i} $\lambda$6678 only shows P Cygni absorption at early
  times.}
\label{fig:vel}
\end{figure}

Since emission lines of He~{\sc i} require relatively high excitation
in the wind compared to H$\alpha$, the ratio of H$\alpha$ to these
He~{\sc i} lines is sensitive to temperature changes
\citep{crowther95,smith95}.  The EW evolution for each of the two
He~{\sc i} lines shown in Figure~\ref{fig:ew} behaves somewhat
differently with time, however.  The EW of He~{\sc i} $\lambda$5876
maintains a constant ratio of about 1/6 of EW(H$\alpha$) within the
uncertainties, whereas the EW of He~{\sc i} $\lambda$6678 is below
this trend for the first $\sim500$ days during and after peak, but
then follows the same trend as the EW of He~{\sc i} $\lambda$5876 for
the remainder of the observations.  This time period when He~{\sc i}
$\lambda$6678 seems to have a weaker EW coincides with a time period
when this line also has a relatively strong P~Cygni absorption
feature, as can be seen in Figure~\ref{fig:he6680}.  The time period
with this deficit is annotated as ``6678 P Cyg'' in
Figure~\ref{fig:ew}, after which the P Cyg absorption disappears.
He~{\sc i} $\lambda$5876 shows blueshifted P~Cygni absorption in its
profile as well (Figure~\ref{fig:he5876}), but in this case the
P~Cygni absorption remains present at all epochs.  Since the
EW(H$\alpha$)/EW(He~{\sc i}) ratios remain roughly constant (except
for the early period with extra P~Cygni absorption in He~{\sc i}
$\lambda$6678), it is likely that the temperature in the line-forming
region of the wind never drops below about 29~kK during the eruption.
In fact, the star gets even hotter at late times (see below).

Unlike He~{\sc i} $\lambda$5876 and $\lambda$6678, the He~{\sc i}
$\lambda$7065 line does change its relative strength with time, being
relatively stronger when the LBV is near the peak of its eruption (see
Figure~\ref{fig:spec}). This may be a density effect, since the
He~{\sc i} $\lambda$7065/$\lambda$5876 flux ratio is sensitive to
density, with relatively stronger $\lambda$7065 emission at higher
densities \citep{an89}.  This implies a higher mass-loss rate in
eruption.

While there are subtle temperature shifts that cause MCA-1B to vary
among the late-WN subtypes, as discussed in the next section, it is clear
that this eruption does not fit the traditional description of an
S~Doradus outburst.  In particular, this is unlike previously studied
cases where an Ofpe/WN9 star was observed to undergo a standard 
S~Doradus LBV eruption, when the temperature became much cooler and the
star became an F-type supergiant, as in the classic
cases of AG~Car in the Milky Way and R127 in the LMC
\citep{ws82,stahl83}.

At the peak of its eruption, its spectral morphology seemed consistent
with the Ofpe/WN9 progenitor in the 1990s \citep{crowther95,smith95}.
This implies that --- in contrast to standard LBVs that show dramatic
temperature variations as they erupt --- {\it MCA-1B appears to have
  brightened by $\sim 1.4$ mag while staying at roughly constant
  temperature}.  Moreover, that temperature was hot (29~kK) as
compared to the usual cool eruptive states of LBVs (7500--8500~K).

If the effective temperature was roughly constant as it brightened,
then MCA-1B's constant temperature requires a significant (factor of
3--4; $\sim 1.4$ mag) increase in bolometric luminosity.  As such, the
outburst of MCA-1B is more like a giant eruption than an S~Doradus
event.  Some giant eruptions do evolve to cooler temperature at peak,
as in the case of light echoes from $\eta$~Car
\citep{rest12,prieto14}.  Other cases remain uncertain; for example,
we have no information about the temperature evolution in P~Cygni's
1600--1650 CE eruption.  There are some known examples of
extragalactic LBV eruptions and SN impostors that show relatively hot
temperatures throughout their evolution.  These include V1 in NGC 2366
\citep{drissen}, HD~5980 in the SMC \citep{drissen,barba95}, GR~290 in
M33 \citep{polcaro16}, SN2000ch \citep{smith11}, and the progenitor
outbursts of SN~2009ip \citep{smith10}, although most of these objects
did not exhibit WN-type spectra at maximum brightness.  In a
decade-long {\it HST} study of bright variables in M51,
\citet{conroy18} detected several bright blue variable stars that
brighten or fade by $\sim$1 mag without a strong colour change (see
their Figure 13).  Thus, MCA-1B seems to be a member of a growing
class of LBV eruptions that do not conform to standard expectations of
cooler peak temperatures.  The fact that MCA-1B was an Ofpe/WN9 star
that underwent an eruption, even though it remained hot, makes an
interesting counterpoint to other LBV eruptions where an Ofpe/WN9 star
erupts but becomes much cooler.  The physical mechanims that control
the temperature evolution (hot or not) during an LBV eruption remain
poorly understood, but might be related to low metallicity or
binarity.

\subsection{The even hotter post-eruption star}

We noted above how little the temperature changed during the eruption
of MCA-1B, at least compared to classical LBV eruptions.  MCA-1B
retained a late-WN spectrum at all times and never became much cooler
as classical S~Dor variables do when they exhibit F-type spectra.  A
closer look, however, reveals subtle temperature variation, shifting
among late-WN subtypes.

This is evident when looking at lines in the blue spectrum,
around H$\beta$ and the blue WR bump, during the long decline for
several years after peak brightness.  This portion of the spectrum is
shown in Figure~\ref{fig:heii}, although note that only some of our
spectral epochs covered the relevant blue wavelengths, and several of
those (not shown in Figure~\ref{fig:heii}) had significant
contamination from the nearby WC star.  In Figure~\ref{fig:heii} we
only show epochs judged to have little contamination from that WC
star.

Although MCA-1B had a late-WN spectrum at all times, the changes to
note in particular concern the relative strengths of He~{\sc i} and
He~{\sc ii} lines, plus H and N lines.  At early times (October 2010),
He~{\sc i} lines like $\lambda$4492 and $\lambda$5016 are quite
strong, but they weaken or disappear at later epochs.  In contrast,
He~{\sc ii} $\lambda$5411 was absent at early times and He~{\sc ii}
$\lambda$4542 was a barely-detected P~Cygni profile, but these lines
become stronger at late times with He~{\sc ii} $\lambda$5411 clearly
in emission.  Notably, the blue WR bump (He~{\sc ii} $\lambda$4686 +
N~{\sc iii} $\lambda$4634) is present at early times in eruption, but
much weaker than H$\beta$.  After about 2012, He~{\sc ii}
$\lambda$4686 becomes as strong as (or stronger than) H$\beta$.

Measured EWs of some of these blue lines are shown in
Figure~\ref{fig:ewhe2}.  While H$\beta$ has a gradual decline
qualitatively like H$\alpha$, it is quite clear that the blue He~{\sc
  i} lines show a dramatic anticorrelation with He~{\sc ii}, with the
He~{\sc ii} lines strengthening as He~{\sc i} lines fade much more
quickly than Balmer lines.  Notice the different behaviour with time
of He~{\sc i} $\lambda$5016 as compared to He~{\sc ii} $\lambda$5411.
These changes require temperature shifts rather than just changes in
mass-loss rate.  Examining spectral classification criteria for WN
stars \citep{crowther95,smith95}, MCA-1B resembles a subtype of WN10
or WN11 at times near and shortly after the peak of the eruption in
2010, and then after 2012 it shifts to a hotter subtype more like WN8
or even WN7.  These changes in WN subtype would correspond roughly to
a shift in temperature from around 29~kK near the peak of the
eruption, to hotter temperatures around 40~kK or more at late times
\citep{hamann95,hamann06}.  MCA-1B in its post-eruption state is
significantly hotter than the S~Dor instability strip for its
luminosity. The precise value of the stellar temperature indicated by
the spectrum requires a more detailed analysis and careful modeling of
the spectrum, but the changes in WN subtype described here are
sufficient to note that the post-eruption star shifted to
significantly higher temperatures than for the 1990s progenitor or the
peak of the eruption.  These changes are extremely unusual for an
LBV-like eruption, and they constitute the first case where an
outbursting massive star retained a WN spectrum throughout its
eruption.

\subsection{Changing outflow speed and radius}

Figures~\ref{fig:ha}, \ref{fig:he6680}, and \ref{fig:he5876} show the
evolution of emission-line profiles for H$\alpha$, He~{\sc i}
$\lambda$6678, and He~{\sc i} $\lambda$5876, respectively.  In each
panel, the top few spectra coincide with times around maximum
luminosity in MCA-1B's eruption, and the lower portion of each panel
traces the late-time quiescent phase.  It is apparent from these
figures that the emission lines were narrower at early times, and
broader at later times when the outburst faded.  In each panel, the
bottom tracing corresponding to the last epoch (black) is overplotted
with the first epoch (orange) at times near peak.  This direct
comparison shows clearly that the lines were narrower near peak
luminosity.  This is most apparent in the He~{\sc i} $\lambda$5876
line, which shows P~Cygni absorption at all epochs, and where the
centre of the P~Cygni trough has clearly shifted to higher blueshifted
velocity at late times.

Figure~\ref{fig:vel} shows the measured FWHM of 
each of these three emission lines as a function of time.
While each line has a slightly different width because they are formed
at different depths in the wind, all three show the same trend of
steadily increasing velocity with time as the brightness faded (the
shaded blue area is a representation of the light curve for
comparison).  Note that at times near peak, He~{\sc i} $\lambda$6678
shows P~Cygni absorption that then goes away, making the overall
contrast in width larger for this line, because the P~Cygni absorption
reduces the emission component's width an additional amount near peak.
(H$\alpha$ shows no P~Cyg absorption at any epoch, and He~{\sc i}
$\lambda$5876 shows P~Cyg absorption at all epochs.)  The wind speed
near peak was around 300--400 km s$^{-1}$, while the lines broadened to
more like 800--900 km s$^{-1}$ at late times when the star faded.  The
H$\alpha$ FWHM was around 500 km s$^{-1}$ for the 1990s progenitor
\citep{smith95}.

Thus, while MCA-1B did not show a major shift to cooler temperatures
at peak brightness, it did exhibit a clear change in wind speed. In
the radiation-driven winds of hot stars, one expects the wind speed to
be roughly proportional to the star's escape speed.  Typically for hot
stars above 21~kK, one expects $v_{\infty}/v_{\rm esc} \approx 2.6$
\citep{lamers95}.  Since the temperature did not drop substantially,
this approximate ratio should hold throughout the eruption.  As the
star's radius changes, the outflow speed should scale roughly as
$v_{\infty} \propto R^{-0.5}$.

First, we compare the pre-eruption star to the properties at the peak
of the eruption.  MCA-1B brightened by 1.4 mag at roughly constant
emitting temperature compared to the 1990s progenitor. The star must
have increased its bolometric luminosity by a factor of 3.0--3.6.
Without any significant change in temperature from the 1990s
progenitor to the eruption peak, this increase in $L$ also requires an
inflation of the star's photospheric radius by a factor of 1.7--1.9.
This larger radius would reduce the star's escape speed and presumably
its wind terminal speed to about 75\% of its pre-outburst value.  This
is in reasonable agreement with the drop in H$\alpha$ FWHM from around
500 km s$^{-1}$ to 400 km s$^{-1}$ (80\%), or the drop in the He~{\sc
  i} $\lambda$6678 width from 420 km s$^{-1}$ to 300 km s$^{-1}$
(70\%).

Next, consider the several years during MCA-1B's post-eruption decline
in brightness, when emission-line widths steadily increased.  The wind
speed roughly doubled, with the H$\alpha$ FWHM rising from around 400
km s$^{-1}$ up to 800--900 km s$^{-1}$ (Figure~\ref{fig:vel}).  As
noted earlier, the velocity of the P~Cygni trough of He~{\sc i}
$\lambda$5876 also increased in speed from about $-330$ km s$^{-1}$ to
roughly $-680$ km s$^{-1}$.  Instead of returning to the temperature
of the 1990s progenitor, MCA-1B became even hotter while fading to
roughly the same visual magnitude.  This implies that the star was
even more compact after the eruption than before, and hence, had an
even higher escape velocity.  The effective temperature after eruption
is uncertain, but may have plausibly been 40~kK or more
\citep{hamann06}, based on the WN8 or WN7 spectral type indicated by
blue emission-line ratios.  If the bolometric luminosity dropped by a
factor of $\sim 3$ from peak to post-eruption quiescence, and the
temperature increased from about 29~kK to 40~kK or more, then we
should expect the wind velocity to increase by a factor of 1.8--1.9
from peak to the post-eruption state, just from the change in stellar
radius.  This is not far from the factor of $\sim2$ increase in wind
speed that is observed.

The slower wind speed at peak luminosity might also help account for
the changes in observed emission-line strengths.  While the line
ratios of strong He~{\sc i} lines (like $\lambda$5876 and
$\lambda$6678) to H$\alpha$ did not change much during the eruption,
the larger EW values at peak brightness mean that all these
emission-line fluxes were stronger at peak.  The EW values were about
2 times higher at peak (Figure~\ref{fig:ew}) when the continuum was
about 3.6 times brighter, indicating a factor of 7--8 increase in
emission-line fluxes at peak as compared to late-time quiescence.  A
slower wind can make emission lines stronger, even with no change in
mass-loss rate, because the emissivity of these lines varies with
electron density as $n_{\rm e}^2$.  If the wind speed at maximum
luminosity slowed to half of the quiescent value, the wind density
would have been 2 times higher, and the lines could have been 4 times
stronger (this assumes that the lines form at radii much larger than
the photospheric radius, which is also changing).  The relatively
strong He~{\sc i} $\lambda$7065 emission in eruption, noted above as a
density effect \citep{an89}, is consistent with a denser eruption
wind.

The larger emission-line strengths at peak may indicate, very roughly,
an increase in the star's mass-loss rate by a modest factor of around
2 during eruption.  A more precise quantitative constraint on the
mass-loss rate and emitting temperature would require a more detailed
model of the spectrum.  This was already done for the Ofpe/WN9
progenitor, yielding a mass-loss rate of roughly $10^{-4}$ $M_{\odot}$
yr$^{-1}$ \citep{smith95}.  This mass-loss rate is already near the
upper limit of what can be achieved with standard line-driven winds
\citep{so06}, and so an increase in both $L$ and the mass-loss rate
above the quiescent value may indicate a transition to a
super-Eddington continuum-driven wind \citep{owocki04}.  Massive stars
already reach appreciable values of $\Gamma = 0.4$ to 0.5 in their
late main sequence \citep{schaller92,sc08}, so an increase in the
bolometric luminosity by a factor of a few would plausibly have driven
MCA-1B to a mildly super-Eddington state.

\begin{figure*}
  \includegraphics[width=4.9in]{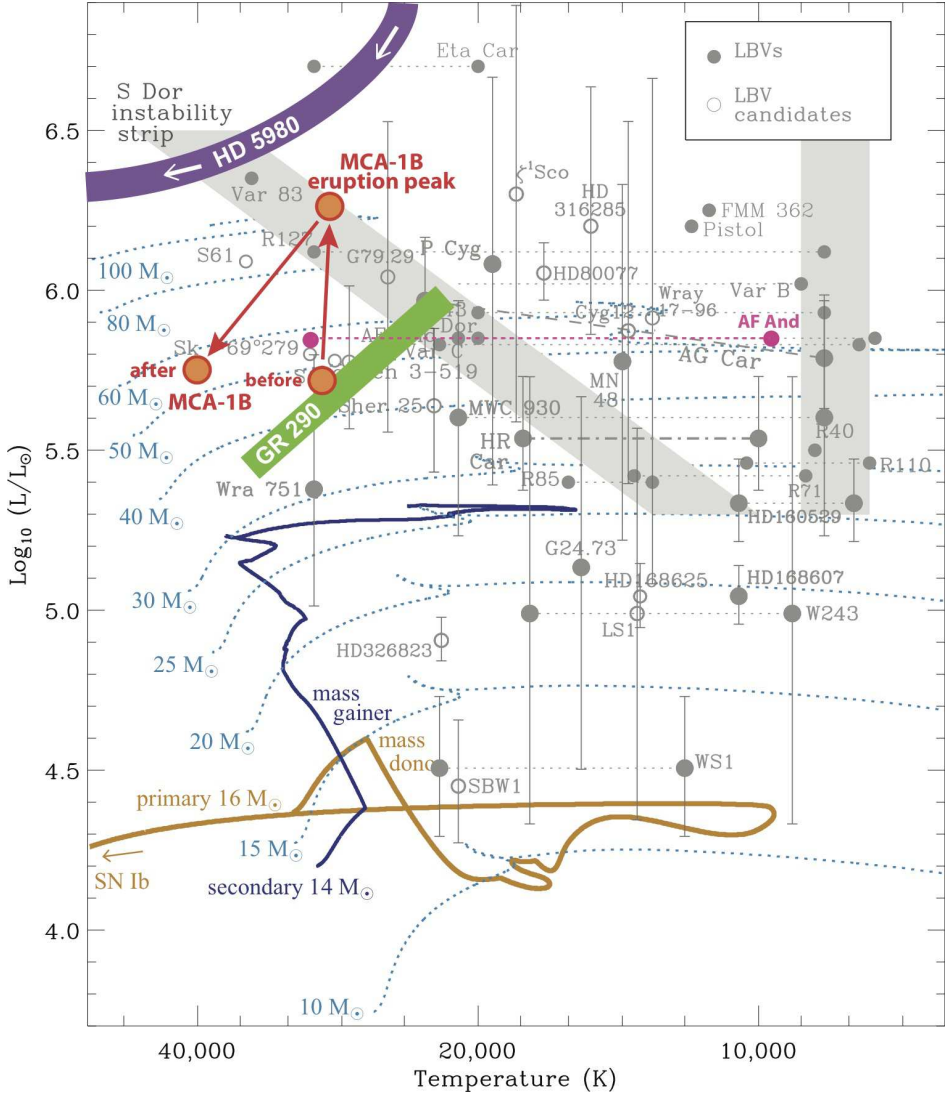}
\caption{MCA-1B in outburst and quiescence (before/after) located
  approximately on the HR diagram (red/orange circles).  These points
  adopt rough temperatures from the WN spectral subtype, and are
  scaled from the progenitor luminosity derived by \citet{smith95} to
  account for contaminated light from unresolved sources. Other LBVs
  are shown for comparison (from \citealt{smith19}), and this figure
  is adapted from that paper.  As in that previous figure,
  stellar evolution models are from \citet{brott11} and
  \citet{lk14}. The purple arc in the upper left represents a range of
  values reported for HD~5980 during its 1994 eruption and in the
  decades afterward \citep{drissen,georgiev11,hillier19}.  The green
  diagonal bar represents a range of values estimated for Romano's
  star (GR~290) from peak to quiescence after its 1990s eruption
  \citep{maryeva19}.  We also show the position of AF~And in M31, as
  recently derived from photometry and spectroscopy by
  \citet{joshi19}, although we note that earlier studies found a
  higher luminosity \citep{szeifert96}.}
\label{fig:hrd}
\end{figure*}

\subsection{On the HR diagram}

It is instructive to compare the properties of MCA-1B at quiescence
and eruption to other LBVs placed on the HR diagram.  Following
\citet{crowther95} and the discussion above we adopt a constant
temperature of about 29~kK for the progenitor and eruption peak, then
transitioning to a hotter temperature of around 40~kK or more in the
years after eruption.  We adopt a quiescent luminosity of
log($L/L_{\odot}$) = 5.72 and a luminosity at the peak of its eruption
in late 2010 of log($L/L_{\odot}$) = 6.28.  While admittedly the
temperature may not be precisely constant before and during eruption
(we have not actually derived a quantitative temperature value at each
epoch of spectra), it clearly does not drop below $\sim10$~kK like
other eruptive LBVs.  With this assumed temperature behaviour, MCA-1B
is shown on the HR diagram in Figure~\ref{fig:hrd}, along with several
other known LBVs (from \citealt{smith19}).

In its hot quiescent state, MCA-1B does not reside on the S~Dor
instability strip, as LBVs are traditionally expected to do; instead,
the progenitor is well below and/or hotter than the strip 
(Fig.~\ref{fig:hrd}).  With revised distances or quantitative
spectroscopic analysis, several other LBVs and LBV candidates are also
located in this area, such as the LBV candidates Hen~3-519 in the
Milky Way and Sk~$-69$~279 and S119 in the LMC, as well as the LBVs
AF~And and Wra~751 (see \citealt{smith19} and references therein).

Interestingly, when MCA-1B brightened in outburst, it did then move
onto the S~Dor instability strip.  This is very unusual behaviour for
LBV-like stars; LBVs generally reside on the S~Dor strip in their hot
quiescence, and they are thought to move horizontally to cool
temperatures.  Those temperature shifts at (assumed) constant
bolometric luminosity are indicated by the dashed horizontal lines for
other LBVs in Figure~\ref{fig:hrd}.  It is very clear that MCA-1B did
not take on a cooler F-type supergiant spectrum in this way.  When its
eruption ended, MCA-1B moved to even hotter temperatures far from the
S~Dor instability strip.

Consider the two contrasting cases of MCA-1B in M33 and the classic
Hubble-Sandage variable AF~And in M31.  Both stars sit in a similar
location on the HR diagram at quiescence.  The $R$-band light curves
of eruptions from these two stars are almost identical in terms of
amplitude, duration, decay rate, and a long period of post-eruption
quiescence, except that AF~And is a bit brighter at all times
(Figure~\ref{fig:phot2}).  The stark difference is that at peak
brightness, MCA-1B retained its hot Ofpe/WN9 spectrum, whereas AF~And
has been seen to cool in its brighter phases to around 10~kK
\citep{jsg81}, as traditionally expected for S~Dor variables.  It was
thought to cool similarly in its most recent eruptions based on
photometric colours \citep{joshi19}, although spectra of these cooler
phases in the most recent eruption have not been published.  Another
difference is that MCA-1B transitioned to even hotter temperatures (a
WN7/8 subtype) after eruption.

This behaviour of MCA-1B is thus far almost unprecedented among
well-studied LBVs.  While we noted above some cases of LBV-like
eruptions that remained hot at peak (SN~2009ip, SN~2000ch, and V1 in
NGC~2366), these are more dramatic cases of SN impostors or pre-SN
outbursts; they were not as hot and did not show WN-type spectra at
their eruption peak, and their properties in their quiescent states
were not as well characterised.  As noted earlier, there are luminous
and blue variable stars in M51 that move vertically in a
colour-magnitude diagram (i.e., they brighten at constant colour;
\citealt{conroy18}).  Perhaps these are something similar, although
they have not been studied spectroscopically yet.  There are,
however, two intriguingly similar cases among spectroscopically
studied LBVs, discussed next.

\begin{figure}
  \includegraphics[width=2.9in]{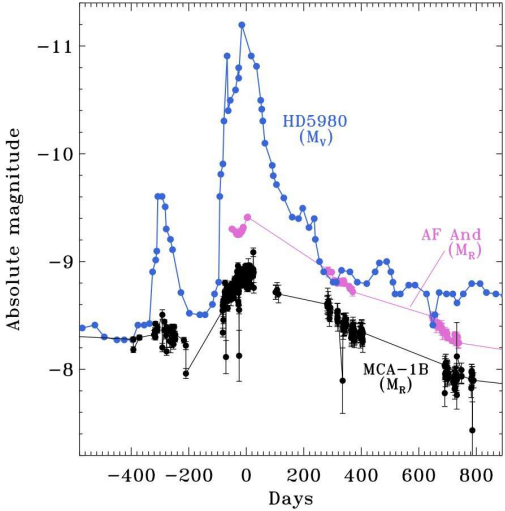}
\caption{The visual absolute magnitude light curve for the 1994
  eruption of HD~5980 in the SMC (shown in blue), converted from
  apparent visual magnitude published by \citet{js97}, as compared to
  the $R$-band light curves of AF~And in pink and MCA-1B in black
  (same as in Figure~\ref{fig:phot2}.)  The spread in magnitudes in
  the original light curve for HD~5980 is larger than shown here
  because of dips due to eclipses; see \citet{js97} for details.
  Here, as in \citet{smith11}, we omit eclipse dips from the visual
  light curve.}
\label{fig:hd5980}
\end{figure}

\subsubsection{HD~5980 in the SMC}

One similar case is the massive LBV system HD~5980 in the SMC, which
is a multiple massive-star system containing an eclipsing binary of
two WN stars, of which the erupting member has retained significant
hydrogen in its atmosphere.  It is the most luminous star in the SMC,
and had a dramatic LBV eruption in 1994; see the recent detailed study
of its post-eruption behaviour by \citet{hillier19}, and previous
reviews and investigations of its evolution during and after eruption
\citep{barba95,moffat98,drissen,gloria94,gloria04,
  gloria14,georgiev11}.  Although it is the most luminous star in the
SMC, it is unexpectedly isolated, being located $\sim$30~pc outside
the massive central cluster of NGC~346, and $\sim$20~pc from any other
known O-type star \citep{st15}.

The light curve of HD~5980's eruption is shown in
Figure~\ref{fig:hd5980}, where it is compared with AF~And and MCA-1B.
Although these are different filters (visual estimates for HD~5980,
and $R$ for MCA-1B and AF~And), both HD~5980 and MCA-1B experienced
little colour change because they remained hot and optical bands are
on the Rayleigh-Jeans tail.  All three eruptions have similar shape
and timescale, but HD~5980 has a larger increase in luminosity.  This
plot even makes it seem as though HD~5980 and MCA-1B both have a
weaker precursor eruption at around $-300$ days relative to the main
peak, although it is difficult to interpret the significance of this
with available data.

The erupting star in HD~5980 has since transitioned to an earlier WN
type after the eruption and the wind speed has increased as the star
became hotter \citep{hillier19}, qualitatively similar to the
behaviour of MCA-1B in its latest phases reported here.  The
approximate post-eruption trajectory of HD~5980 on the HR diagram is
indicated by the purple arc in Figure~\ref{fig:hrd}.  Unlike MCA-1B,
the eruption spectrum of HD~5980 was described as B-type (B1.5 Ia+)
with a cooler temperature of roughly 23~kK \citep{drissen,barba95}.
The WR features significantly weakened or disappeared in eruption
\citep{barba95,gloria04}, although a WN11-like spectrum during
eruption was also reported \citep{moffat98}.  There are some
indications that He~{\sc ii} $\lambda$4686 was present in the erupting
star, not the companion \citep{hillier19}.  In any case, HD~5980 was
similar to MCA-1B in the sense that it clearly did not transition to a
very cool temperature (F-type, $\sim8000$~K) in outburst.  In terms of
its overall observed properties, the 2010 eruption of MCA-1B seems
quite similar to the 1994 eruption of HD~5980 (although MCA-1B is less
luminous).  Both are shown on the HR diagram in Fig.~\ref{fig:hrd}.

The interpretation of HD~5980 is complicated by the fact that it is
known to be a member of a short-period (19.3 day) eclipsing binary
containing a massive WNE companion \citep{gloria14}.  Of course, we
cannot rule out the possibility that MCA-1B shares similar
complications, and this may be an important clue to understanding this
peculiar subset of LBV behaviour in binaries.  It could be interesting
to monitor MCA-1B to search for radial-velocity variations. (Our
lower resolution spectra are not adequate for this, but some of our
higher resolution spectra are.  Note that the line profiles in Figures
7, 8, and 9 have been shifted to emphasise line profile width and
shape.  A future paper will analyze possible radial-velocity
variations after we obtain more data and subtract contaminating
emission from the WC star MCA-1.)

\subsubsection{Romano's star (GR~290) in M33}

Romano's star \citep{romano78}, also called GR~290 or v532, is a
luminous and peculiar LBV that is sometimes referred to as an LBV
candidate because it does not quite fit expectations for typical LBVs
--- namely, it stays hot and shows little or no colour variation
during eruptions, like MCA-1B.  This LBV has been studied in detail;
see the recent overview by \citet{maryeva19} and several other studies
\citep{romano78,polcaro03,polcaro16,kurtev01,fabrika05,viotti06,viotti07}.

GR~290 also resides in M33, and like MCA-1B is also found in its
somewhat remote outer parts ($\sim 4$ kpc from the centre), but on the
eastern side of the galaxy.  Like most LBVs, it is also relatively
isolated, located about 125~pc from the nearest OB association OB89
\citep{maryeva19}.

GR~290 had a major eruptive episode that peaked in 1994, although its
photometric evolution was much slower than that of MCA-1B and HD~5980.
GR~290 shows a late WN-type spectrum at most epochs, although at the
peak of its eruption in 1994 it showed a mid to late B-type spectrum
\citep{polcaro16}, somewhat cooler than MCA-1B and more like HD~5980.
As it faded for many years after its eruption, it slowly cascaded
through late WN types, from WN10/11 to WN8, very much as MCA-1B did.
The post-eruption trajectory of GR~290, adapted from
\citet{maryeva19}, is shown by the green diagonal bar on
Figure~\ref{fig:hrd}. (Note that \citealt{polcaro16} also document its
post-outburst evolution, but have luminosity values shifted about 0.1
dex higher.)  This behaviour resembles that of MCA-1B, and seems like
a less luminous version of HD~5980.  It is not yet known if GR~290 is
a member of a binary system.  GR~290 has been discussed as an LBV or a
post-LBV object in transition to the WN phase, although such
speculation about its age and evolution were based on single-star
evolutionary models that might be inapplicable to these objects.

\subsection{A subclass of WN-type LBVs}

The class of massive stars known as LBVs is already a hodgepodge of
irregular and eruptive variables that includes traditional S~Dor
variables, $\eta$~Car variables or giant eruptions, SN impostors,
$\alpha$ Cygni variables, P~Cygni stars, and so on \citep{conti84}.
Given the distinct similarities between MCA-1B, HD~5980, and GR~290,
it seems worthwhile to proliferate yet another subset of LBVs that is
distinct from normal S~Dor variables.  While these three stars provide
a few well-studied examples, there may be many more; we noted earlier
that an unbiased study of variables in M51 found many luminous blue
stars that show little colour change as they brighten and fade
\citep{conroy18}.  Some key observed characteristics shared by MCA-1B,
HD~5980, and GR~290, which set them apart from other LBVs, are as
follows.

(1) In their fainter quiescent states, they have late WN-type spectra
(WN7--WN9).  Although some other S~Dor-type LBVs have Ofpe/WN9-type
spectra in quiescence, the WN-type LBVs reside well off the S~Dor
instability strip (below and to hotter temperatures).

(2) When experiencing significant eruptive episodes where they
brighten by roughly 1 mag or more, they do not shift to much cooler
temperatures (8--10~kK) as S~Dor variables do.  Instead, the WN-type
LBVs stay well above 20~kK at all times, exhibiting late WN-type
spectra at most times (WN10--WN11), with some showing B-type spectra
at their coolest and brightest epochs.

(3) With their significant brightening and with little or no
colour-temperature shift, these stars experience a substantial (factor
of several) increase in bolometric luminosity during outburst.  They
do not, however, develop very cool temperatures as some LBV giant
eruptions do (as in the case of the giant eruption of $\eta$~Car;
\citealt{rest12}).

(4) Although they have WN-type spectra, they have retained some amount
of hydrogen in their envelopes that may, nevertheless, indicate a
lower H/He abundance ratio than other LBVs.

(5)  Somewhat more speculatively, these objects seem to prefer
relatively low metallicity, with HD~5980 residing in the SMC, and both
MCA-1B and GR~290 in the low-metallicity outskirts of M33.

In addition to the possibility that this behaviour occurs in binary
systems (certainly the case for HD~5980), two key issues stand out in
the context of these unusual LBV eruptions.  One is the relatively 
low-metallicity environments, where wind mass-loss may generally be
weaker, and where the potential role of the Fe opacity bump in
triggering the LBV instability \citep{jiang18} may be diminished or
may have different consequences for the amount of envelope inflation.
It is quite likely that the lower metallicity may play an important
role in the lack of observed temperature shifts in eruption.  This may
also significantly influence their evolution and instability, although
there are still few examples of well-studied LBVs at low metallicity.
Another issue is the possibility that such eruptive events may play an
important role in the post-LBV transition to a WR star.  MCA-1B and
HD~5980 suggest that, at low metallicity at least, the eruptive LBV
instability may persist into the early stages of the WR phase of
evolution.

Lastly, an LBV-like eruption where the star remains hot and retains a
WN-type spectrum at all times has possible links to certain peculiar
SNe. Specifically, Type~Ibn SNe represent a class of explosions that
encounter strong interaction with circumstellar material (CSM), as in
SNe~IIn, but where the CSM is H-poor, causing an optical spectrum
dominated by He~{\sc i} lines \citep{pastorello08,hoss17,smith17b}.
The best-studied case so far was SN~2006jc, which had a pre-SN
outburst detected 2~yr before the SN, with a He-rich CSM shell
expanding at around 1000 km s$^{-1}$.  This led to suggestions of a
progenitor that had an LBV-like eruption, but in a hydrogen-deficient
WR star \citep{pastorello07,foley07,smith08}.\footnote{Note, however,
  that evidence suggests not all SNe~Ibn necessarily come from massive
  WR stars \citep{hoss19,sanders13}.} Cases have been seen with a
range of H/He line strengths, suggesting that explosions occur in a
continuum of stages caught between the LBV and WR phases
\citep{smith12,pastorello15}. MCA-1B, GR~290, and HD~5980 are
therefore valuable as potential analogs for the progenitor systems of
such SNe~Ibn, provided that they erupt shortly before death.

\section{SUMMARY}

We present the discovery of a new LBV-like star in M33, indicated by a
nonperiodic brightening in 2010, and we present long-term follow-up
photometry and spectroscopy.  The progenitor star, named MCA-1B, was
already known to be an Ofpe/WN9 star in the 1990s, and had been
studied in detail \citep{willis92,smith95,bianchi04}.  At that time,
\citet{smith95} proposed that it was a dormant LBV, which is now
confirmed.

The 2010--2011 eruption brightened by 1.4 mag, but has since faded,
returning approximately to its pre-outburst brightness.  The evolution
of the light curve was very similar to the classic LBV star AF~And in
M31 \citep{joshi19}, and to some other well-studied LBVs like HR~Car.
It is located in a small star cluster in M33's remote western
outskirts where the metallicity is roughly between that of the LMC and 
SMC, and we discussed the influence that these neighbouring
stars (especially the WC star MCA-1) exert on the ground-based
photometry and spectra.  Observations require that the star increased
its bolometric luminosity by almost a factor of 4 during the eruption,
and increased its radius by about a factor of 2.  MCA-1B may have
reached or exceeded its Eddington limit during outburst.

Overall, the eruption of MCA-1B appears to show several of the
hallmarks of an LBV giant eruption, with an increase in bolometric
luminosity, an inflation of the photospheric radius, a slowing of the
wind speed, an Ofpe/WN9 spectrum in its quiescent state, and possibly
a modest increase in the mass-loss rate during eruption.
Additionally, the overall appearance of the light curve is nearly
identical to that of an eruption of the classic LBV star AF~And in 
M31 (MCA-1B is just $\sim 0.3$ mag fainter).

It is therefore quite interesting and surprising that this eruption of
MCA-1B did {\it not} exhibit one of the most commonly cited properties
of LBV eruptions, which is a shift to cool apparent temperatures
(8000--10,000~K) at peak brightness.  Instead, MCA-1B retained its hot
WN spectral type throughout its eruption.  This behaviour is almost
unprecedented among well-studied LBVs, with the closest analogues
being HD~5980 in the SMC and GR~290 in M33.  It is tempting to
speculate that this lack of a temperature shift might be related to
its relatively low metallicity, suggested by its remote location in
the outskirts of M33.  Like HD~5980, it may also be related to the
progenitor residing in a short-period binary system, although
information about possible binarity is not yet available for MCA-1B.
In any case, the LBV-like eruption of a WN star is potentially
important for understanding the progenitors of SNe~Ibn, which based on
their H-poor circumstellar shells, are inferred to have WR progenitors
that erupted like LBVs.  Moreover, it has been suggested that SN~Ibn
progenitors may have been caught in a brief post-LBV/WR transition
when they erupted and then exploded, as inferred for MCA-1B, GR~290,
and HD~5980.

We noted that there are some possible analogues that brighten without
an accompanying colour change in a recent photometric study of M51
\citep{conroy18}, although these cases lack follow-up spectroscopy.
We also noted a few SN impostors that exhibit hot temperatures at peak
brightness (SN~2000ch, SN~2009ip, and V1 in NGC~2366), but these are
more extreme phenomena.  The location in the outskirts of M33 is
intriguing in this regard, because it is very reminiscent of the
isolated environment of SN~2009ip \citep{smith+16}, which was also
thought to be a massive LBV before death
\citep{smith10,foley11,mauerhan13,mauerhan14,graham14,graham17,reilly17}.

In studies of the variability of LBVs, it is often taken for granted
that they shift to cool temperatures at peak, while maintaining
constant bolometric luminosity \citep{hd94,hds99}.  As for HD~5980 and
GR~290, MCA-1B suggests that this is not a safe assumption without
corroborating evidence from spectra or colours, even if the light
curve otherwise resembles a normal S~Dor phase.  The fraction of LBVs
that lack a significant colour change and may be WN-type LBVs is not
yet quantified.  This, therefore, underscores the diversity in the
class of LBVs, and that some past definitions of LBVs might have been
too narrow.  If MCA-1B were not regarded as a true LBV, then it would
require the invention of yet another new subclass of luminous, blue,
eruptive variables.  The original intent of the term ``LBV''
\citep{conti84} was to group all such observed phenomena together.

\section*{Acknowledgements}

We thank Dovi Poznanski and Eran Ofek for assistance with the PTF
photometry, and Yi Cao, Brad Cenko, Ryan Foley, and Shri Kulkarni for
assistance with both photometric and spectroscopic observations
obtained through PTF.  We are grateful to Yogesh Joshi for sending a
table of photometry for AF~And prior to publication, and we thank Paul
Crowther for providing spectra of the Ofpe/WN9 comparison stars in
Figure 5.  Support for N.S. was provided by NSF award AST-1515559, and
by the National Aeronautics and Space Administration (NASA) through
{\it HST} grant AR-14316 from the Space Telescope Science Institute,
which is operated by AURA, Inc., under NASA contract NAS5-26555.
A.V.F.'s research has been generously supported by the TABASGO
Foundation, the Christopher R. Redlich Fund, and the Miller Institute
for Basic Research in Science (U.C. Berkeley).

This paper is based in part on observations obtained with the Samuel
Oschin Telescope and the 60 inch Telescope at the Palomar Observatory
as part of the Palomar Transient Factory project, a scientific
collaboration between the California Institute of Technology, Columbia
University, Las Cumbres Observatory, the Lawrence Berkeley National
Laboratory, the National Energy Research Scientific Computing Center,
the University of Oxford, and the Weizmann Institute of Science.
Observations using Steward Observatory facilities were obtained as
part of the observing program AZTEC: Arizona Transient Exploration and
Characterization, which receives support from NSF grant AST-1515559.
Some of the data presented herein were obtained at the W. M. Keck
Observatory, which is operated as a scientific partnership among the
California Institute of Technology, the University of California, and
NASA; the observatory was made possible by the generous financial
support of the W. M. Keck Foundation. Research at Lick Observatory is
partially supported by a generous gift from Google. Some observations
reported here were obtained at the MMT Observatory, a joint facility
of the University of Arizona and the Smithsonian Institution.  This
paper uses data taken with the MODS spectrographs built with funding
from NSF grant AST-9987045 and the NSF Telescope System
Instrumentation Program (TSIP), with additional funds from the Ohio
Board of Regents and the Ohio State University Office of Research. We
are grateful for the assistance of the staffs at the various
observatories where data were obtained.

Facilities: HST (WFPC2), Keck:I (LRIS), Keck II (DEIMOS), LBT (MODS),
MMT (Bluechannel), Lick: 3m (Kast), PO:1.2m, 1.5m, SO: Bok (B\&C,
SPOL), SO:Kuiper (Mont4K) SO:Super-LOTIS

\end{document}